\documentclass[%
 reprint,
preprintnumbers,
 amsmath,amssymb,
 aps,
floatfix,
]{revtex4-2}

\usepackage{graphicx}
\usepackage{dcolumn}
\usepackage{bm}
\usepackage{siunitx}
\usepackage{color}      
\usepackage{amsbsy}     
\usepackage{amsfonts}       
\usepackage{amsmath}        
\usepackage{amssymb}        
\usepackage{hyperref}
\usepackage{cleveref}
\usepackage{xcolor}
\usepackage{slashed}
\usepackage{slashed}
\usepackage{graphicx}
\usepackage{physics}
\usepackage{nicefrac}
\usepackage{subcaption}
\usepackage{tikz}
\usepackage[compat=1.1.0]{tikz-feynman}

\DeclareMathOperator\arctanh{atanh}

\DeclareSIUnit \parsec {pc}

\crefformat{section}{Sec.~#2#1#3}
\Crefformat{section}{Sec.~#2#1#3}

\crefformat{appendix}{App.~#2#1#3}
\Crefformat{appendix}{App.~#2#1#3}

\crefformat{table}{Tab.~#2#1#3}
\Crefformat{table}{Table~#2#1#3}

\crefformat{equation}{Eq.~(#2#1#3)}
\Crefformat{equation}{Eq.~(#2#1#3)}
\crefrangeformat{equation}{Eqs.~(#3#1#4) to~(#5#2#6)}
\crefmultiformat{equation}{Eqs.~(#2#1#3)}{ and~(#2#1#3)}{, (#2#1#3)}{ and~(#2#1#3)}

\crefformat{figure}{Fig.~#2#1#3}
\crefrangeformat{figure}{Figs.~(#3#1#4) to~(#5#2#6)}
\crefmultiformat{figure}{Figs.~(#2#1#3)}{ and~(#2#1#3)}{, (#2#1#3)}{ and~(#2#1#3)}

\tikzfeynmanset{
     anti ghost/.style={
        /tikz/draw=none,
        /tikz/decoration={name=none},
        /tikz/postaction={
          /tikz/draw,
          /tikz/dotted,
          /tikz/thick,
          /tikzfeynman/with reversed arrow=0.5,
        },
     },
}
\tikzfeynmanset{
    every ghost@@/.style={
        /tikz/draw=none,
        /tikz/decoration={name=none},
        /tikz/postaction={
            /tikz/draw,
            /tikz/dotted,
            /tikz/thick,
            /tikzfeynman/with arrow=0.5
    },
  },
}

\begin{document}

\preprint{MS-TP-22-16}

\title{Radiative corrections to stop-antistop annihilation into gluons and light quarks}

\author{M.~Klasen}
 \email{michael.klasen@uni-muenster.de}
 \affiliation{
	Institut f\"ur Theoretische Physik, Westf\"alische Wilhelms-Universit\"at M\"unster, Wilhelm-Klemm-Stra{\ss}e 9, D-48149 M\"unster, Germany
  }
\affiliation{School of Physics, The University of New South Wales, Sydney NSW 2052, Australia}

\author{K.~Kova\v{r}\'ik}
 \email{karol.kovarik@uni-muenster.de}
 \affiliation{
	Institut f\"ur Theoretische Physik, Westf\"alische Wilhelms-Universit\"at M\"unster, Wilhelm-Klemm-Stra{\ss}e 9, D-48149 M\"unster, Germany
  }
 \author{L.P.~Wiggering}
 \email{luca.wiggering@uni-muenster.de}
 \affiliation{
	Institut f\"ur Theoretische Physik, Westf\"alische Wilhelms-Universit\"at M\"unster, Wilhelm-Klemm-Stra{\ss}e 9, D-48149 M\"unster, Germany
  }

\date{\today}

\begin{abstract}
We present the full one-loop SUSY-QCD corrections to stop-antistop annihilation into gluons and light quarks within the Minimal Supersymmetric Standard Model including Sommerfeld enhancement effects from the exchange of multiple gluons between the incoming particles. These corrections are important as stop (co)annihilation becomes the dominant contribution to the relic density for scenarios with a small mass difference between the neutralino and the stop which are less constrained by current LHC searches and consistent with the observation of a 125 GeV SM-like Higgs boson. We discuss important technical details of our one-loop, real emission, and resummation calculations where we pay particular attention to the cancellation of infrared divergences and the associated application of the dipole formalism for massive initial scalars. The corrections have been implemented in the dark matter precision tool \texttt{DM@NLO} which allows us to study numerically the impact of these corrections on the annihilation cross section. We find that for the chosen reference scenario the dominant correction comes from the Sommerfeld effect and that the pure NLO correction is below $\SI{3}{\percent}$. The inclusion of these radiative corrections is still large enough to decrease the relic density by more than $\SI{10}{\percent}$ and shift the cosmologically preferred parameter region by a few GeV relative to the standard \texttt{MicrOMEGAs}  result. Therefore, the inclusion of these corrections is mandatory if the experimental errors are taken as upper and lower bounds of the theory value.
\end{abstract}

\maketitle


\section{\label{sec:intro} Introduction}
There is compelling evidence from astrophysical observations that there is a yet unknown type of matter called dark matter (DM) which does not interact electromagnetically but manifests itself through its gravitational effects on baryonic matter \cite{Freese:2017idy}. The most promising candidate for dark matter is a weakly interacting massive particle (WIMP) as it is consistent with structure formation due to its non-relativistic velocity and naturally leads via the freeze-out mechanism to the correct relic density of cold dark matter (CDM) 
\begin{align}
    \Omega_{\rm CDM}h^2 ~=~ 0.120 \pm 0.001
    \label{eq:omh2}
\end{align}
as determined by the Planck satellite within the $\Lambda\text{CDM}$ model \cite{Planck:2018vyg}. The indicated uncertainty corresponds to the $1\sigma$ interval, and $h$ stands for the present Hubble expansion rate $H_0$ in units of $\SI{100}{\kilo\meter\per\second\per\mega\parsec}$. 

As the Standard Model (SM) does not accommodate a suitable DM candidate there is the need for physics beyond the SM. A widely studied extension is the R-symmetric Minimal Supersymmetric Standard Model (MSSM) \cite{Nilles:1983ge} as it contains not only an appropriate WIMP candidate in the form of the lightest neutralino $\tilde{\chi}^0_1$, but also offers a solution to the hierarchy problem and allows for the unification of gauge couplings at high energies. In order to make a theoretical prediction for the relic density of the neutralino under the assumption of the freeze-out scenario, one has to solve the Boltzmann equation 
\begin{equation}
    \dv{n_\chi }{t}= -3 H n_\chi - \langle \sigma_{\text{eff}} v\rangle \left(n^2_\chi-(n_\chi^{\text{eq}})^2\right)
        \label{eq:boltzmann}
\end{equation}
for the DM number density $n_\chi$ where $n_\chi^{\text{eq}}$ denotes the density in chemical equilibrium and $H$ the Hubble rate \cite{Gondolo:1990dk,Edsjo:1997bg}. Today's neutralino relic density is then given by
\begin{equation}
    \Omega_\chi = \frac{m_\chi n^0_\chi}{\rho_{\text{c}}} \sim \frac{1}{ \langle \sigma_{\text{eff}} v\rangle}
\end{equation}
where $n^0_\chi$ denotes the present value for the number density, $m_\chi$ the DM mass  and $\rho_c$ today's critical density. {The number density equation in \cref{eq:boltzmann} is only an all order expression in the zero temperature limit since the phase space distribution functions of the SM particles are no longer exponentially suppressed by energy conservation for more than two particles in the initial or final state. This in principle forbids the usage of Maxwell-Boltzmann statistics and the neglect of Bose enhancement and Fermi blocking factors for $2 \to 3$ processes appearing at the one-loop level in the collision term. However, in Ref. \cite{Beneke:2014gla} the additional thermal corrections  where found to be suppressed by a factor $T_F/m_\chi\ll 1$ compared to zero temperature $\order{\alpha_s}$ corrections with $T_F$ being the freeze-out temperature. The thermal corrections are therefore negligible at the current level of experimental precision of the dark matter relic density justifying the zero-temperature approach.} The thermally averaged effective cross section 
\begin{equation}
    \langle \sigma_{\text{eff}} v\rangle = \sum_{i,j} \langle\sigma_{ij} v\rangle \frac{n^{\text{eq}}_i}{n_\chi^{\text{eq}}} \frac{n^{\text{eq}}_j}{n_\chi^{\text{eq}}} 
    \label{eq:eff_XSec}
\end{equation}
involves a sum over all supersymmetric particles with odd $R$-parity where $\sigma_{ij}$ corresponds to the cross section for the annihilation of $i$ and $j$ into all possible SM particles. For the following analysis it is important to recall that the ratio $\nicefrac{n^{\text{eq}}_i}{n_\chi^{\text{eq}}}$ is Boltzmann suppressed
\begin{equation}
    \frac{n^{\text{eq}}_i}{n_\chi^{\text{eq}}} \sim \exp(-\frac{m_i-m_\chi}{T}) 
    \label{eq:neq}
\end{equation}
with $T$ being the temperature at time $t$. A direct consequence of \cref{eq:neq} is that besides neutralino annihilation only those processes involving other particles from the odd sector in the initial-state with a small mass difference to the DM candidate can contribute significantly to $\langle \sigma_{\text{eff}} v\rangle$. Especially for large neutralino masses, the neutralino annihilation cross section alone is for many scenarios in the MSSM too small to be consistent with the measured relic density. Therefore, the neutralino cross section needs to be enhanced by some mechanism which could be colored (co)annihilation. 

In this paper, we focus on the case where the mass of the lightest stop is very close to the one of the neutralino so that stop-antistop annihilation and stop pair-annihilation become the dominant contribution to the effective cross section, and thus the relic density. This mass hierarchy is not an unnatural assumption since the tree-level mass of the lightest Higgs boson in the MSSM is bounded from above by $m_{Z^0} |\cos2\beta|$ which requires large quantum corrections to be consistent with the observation of a SM-like $\SI{125}{\giga\electronvolt}$ Higgs boson \cite{ATLAS:2012yve,CMS:2012qbp}. The dominant contribution to the Higgs mass comes from the stop sector where a large trilinear coupling $A_t$ is needed in order for these corrections to be large enough, further indicating a large mass splitting between $m_{\tilde{t}_1}$ and $m_{\tilde{t}_1}$ \cite{Arbey:2012bp}. The mass splitting is enhanced further through the fact that the off-diagonal entries in the sfermion mixing matrix are proportional to the associated masses of the SM partners, indicating a rather light $\tilde{t}_1$. 

The very small experimental uncertainty of the relic density in \cref{eq:omh2} requires the inclusion of radiative corrections to the annihilation cross section so that the theoretical precision matches the experimental one. However, public tools for the calculation of the relic density such as \texttt{DarkSUSY} \cite{Gondolo:2004sc} and \texttt{MicrOMEGAs} \cite{Belanger:2001fz,Belanger:2006is,Barducci:2016pcb} only take into account the tree-level cross section with effective couplings that capture certain higher order effects.

The importance of higher-order SUSY-QCD corrections to the relic density has been shown for gaugino pair-annihilation into quarks \cite{Herrmann:2007ku, Herrmann:2009wk, Herrmann:2009mp, Herrmann:2014kma}, gaugino-squark coannihilation into final states with a quark \cite{Freitas:2007sa, Harz:2012fz, Harz:2014tma}, squark-antisquark annihilation into electroweak final states \cite{Harz:2014gaa}, squark pair-annihilation into quarks \cite{Schmiemann:2019czm} and stau-antistau annihilation into heavy quarks \cite{Branahl:2019yot}. Furthermore, the reduction of theoretical uncertainties from scheme and scale variations have been examined systematically \cite{Harz:2016dql,Branahl:2019yot}. Electroweak corrections to neutralino annihilation have been computed in \cite{Boudjema:2005hb, Baro:2007em, Baro:2009na}. {It should be noted that the previous non-exhaustive list focuses only on one-loop corrections for relic density calculations. However, higher-order corrections in other contexts can also play an important role. The supersymmetric one-loop corrections in the strong coupling to the elastic neutralino-nucleon cross section relevant for direct detection were for example examined in Ref. \cite{Klasen:2016qyz}, and one-loop EW corrections to Wino dark matter annihilation for indirect detection signals were computed in Ref. \cite{Hryczuk:2011vi}.}

Based on these findings we present in this paper corrections of $\order{\alpha_s}$ including Sommerfeld enhancement effects to the processes
\begin{subequations}
\label{eq:processes}
\begin{align}
    \tilde{t}_1 \tilde{t}^\ast_1 ~&\longrightarrow~ g g \,, \label{eq:process_gg} \\
    \tilde{t}_1 \tilde{t}^\ast_1 ~&\longrightarrow~ q \bar{q} \,, \label{eq:process_qqbar} 
\end{align}
\end{subequations}
with the effectively massless quarks $q \in \{u,d,c,s\}$. These two processes are separate at tree level but have to be merged into one at NLO accuracy in order to obtain an infrared safe cross section. 

The paper is organized as follows: in \cref{sec:pheno} we present the color decomposed leading order cross section and discuss the phenomenological relevance of stop-antistop annihilation on the basis of a viable reference scenario. \Cref{sec:compDetails} covers details on the calculation of the virtual and real corrections, followed by the Sommerfeld resummation. In \cref{sec:results}, we discuss the impact of the corrections on the corresponding cross section as well as the relic density for the chosen reference scenario. We conclude in \cref{sec:conclusion}.

\section{\label{sec:pheno}Phenomenology of squark-antisquark annihilation}
To prepare for the subsequent discussion of the higher order corrections and to clarify the notation, we start with the analytic computation of the tree-level cross section and discuss the phenomenology of the processes in \cref{eq:processes} in the context of the neutralino relic density.

\subsection{\label{sec:LO-XSec}Leading order cross section}
\begin{figure*}[t!]
\centering
\begin{subfigure}[c]{\textwidth}		
\begin{tikzpicture}[scale =0.85]
\begin{feynman}
 \vertex (v0) at (0.0,3.49){\({\tilde{t}_{i,t}}\)};
  \vertex (v1) at (1.4,2.33);
  \vertex (v2) at (0.0,1.16){\({\tilde{t}_{j,s}^{\ast}}\)};
  \vertex (v3) at (4.65,3.49){\({g^{\mu}_a}\)} ;
  \vertex (v4) at (3.26,2.33);
  \vertex (v5) at (4.65,1.16){\({g^{\nu}_b}\)};
\diagram* {
(v0) -- [charged scalar,momentum={\(p_{a}\)}] (v1),(v2) -- [anti charged scalar,momentum'={\(p_{b}\)}] (v1),(v3) -- [gluon, reversed momentum'={\(k_{1}\)}] (v4),(v5) -- [gluon,reversed momentum={\(k_{2}\)}] (v4),(v1) -- [gluon] (v4),};
\end{feynman}
\end{tikzpicture} \hspace{2.5cm}
\begin{tikzpicture}[scale =0.85]
\begin{feynman}
  \vertex (v0) at (0.0,3.49);
  \vertex (v1) at (2.33,3.26);
  \vertex (v2) at (0.0,1.16);
  \vertex (v3) at (2.33,1.4);
  \vertex (v4) at (4.65,3.49);
  \vertex (v5) at (4.65,1.16);
\diagram* {
(v0) -- [charged scalar] (v1),(v2) -- [anti charged scalar] (v3),(v4) -- [gluon] (v1),(v5) -- [gluon] (v3),(v1) -- [charged scalar] (v3),};
\end{feynman}
\end{tikzpicture}

\vspace{.4cm}
\begin{tikzpicture}[scale =0.85]
\begin{feynman}
  \vertex (v0) at (0.0,3.49);
  \vertex (v1) at (2.33,3.26);
  \vertex (v2) at (0.0,1.16);
  \vertex (v3) at (2.33,1.4);
  \vertex (v4) at (4.65,3.49);
  \vertex (v5) at (4.65,1.16);
\diagram* {
(v0) -- [charged scalar] (v1),(v2) -- [anti charged scalar] (v3),(v4) -- [gluon] (v3),(v5) -- [gluon] (v1),(v1) -- [charged scalar] (v3),};
\end{feynman}
\end{tikzpicture} \hspace{3.2cm}
\begin{tikzpicture}[scale =0.85]
\begin{feynman}
  \vertex (v0) at (0.0,3.49);
  \vertex (v1) at (2.33,2.33);
  \vertex (v2) at (0.0,1.16);
  \vertex (v3) at (4.65,3.49);
  \vertex (v4) at (4.65,1.16);
\diagram* {
(v0) -- [charged scalar] (v1),(v2) -- [anti charged scalar] (v1),(v3) -- [gluon] (v1),(v4) -- [gluon] (v1),};
\end{feynman}
\end{tikzpicture}
\subcaption{Graphs for the annihilation into two gluons given by the amplitude $\mathcal{M}_{gg}$.}
\end{subfigure}
\begin{subfigure}[c]{0.45\textwidth}
\begin{tikzpicture}[scale =0.85]
\begin{feynman}
 \vertex (v0) at (0.0,3.49){\({\tilde{t}_{i,t}}\)};
  \vertex (v1) at (1.4,2.33);
  \vertex (v2) at (0.0,1.16){\({\tilde{t}_{j,s}^{\ast}}\)};
  \vertex (v3) at (4.65,3.49){\({q_r}\)} ;
  \vertex (v4) at (3.26,2.33);
  \vertex (v5) at (4.65,1.16){\({\bar{q}_u}\)};
\diagram* {
(v0) -- [charged scalar,momentum={\(p_{a}\)}] (v1),(v2) -- [anti charged scalar,momentum'={\(p_{b}\)}] (v1),(v3) -- [anti fermion, reversed momentum'={\(k_{1}\)}] (v4),(v5) -- [fermion,reversed momentum={\(k_{2}\)}] (v4),(v1) -- [gluon] (v4),};
\end{feynman}
\end{tikzpicture} 
\subcaption{Graph for the annihilation into a massless quark-antiquark pair given by the amplitude $\mathcal{M}_{q\bar{q}}$.}
\end{subfigure}
\begin{subfigure}[c]{0.45\textwidth}
	\begin{tikzpicture}[scale =0.85]
\begin{feynman}
 \vertex (v0) at (0.0,3.49){\({\tilde{t}_{i,t}}\)};
  \vertex (v1) at (1.4,2.33);
  \vertex (v2) at (0.0,1.16){\({\tilde{t}_{j,s}^{\ast}}\)};
  \vertex (v3) at (4.65,3.49){\({c_a}\)} ;
  \vertex (v4) at (3.26,2.33);
  \vertex (v5) at (4.65,1.16){\({\bar{c}_b}\)} ;
\diagram* {
(v0) -- [charged scalar,momentum={\(p_{a}\)}] (v1),(v2) -- [anti charged scalar,momentum'={\(p_{b}\)}] (v1),(v3) -- [anti ghost, reversed momentum'={\(k_{1}\)}] (v4),(v5) -- [ghost,reversed momentum={\(k_{2}\)}] (v4),(v1) -- [gluon] (v4),};
\end{feynman}
\end{tikzpicture}
\subcaption{Graph for the annihilation into a ghost-antighost pair given by the ampltidue $\mathcal{S}_1^{\text{Tree}}$. The amplitude for $\mathcal{S}_2^{\text{Tree}}$ is obtained by reversing the ghost flow.}
\end{subfigure}

\caption{Tree-level Feynman diagrams associated with the annihilation of a stop-antistop pair into gluons and quarks. Four-momenta ($p_a,p_b,k_1,k_2$), sfermion indices $(i,j)$, colors ($s$,$t$,$a$,$b$,$r$,$u$) and Lorentz indices ($\mu,\nu$) are explicitly labeled in the respective first diagrams.}
\label{fig:TreeLevelDiags}
\end{figure*}
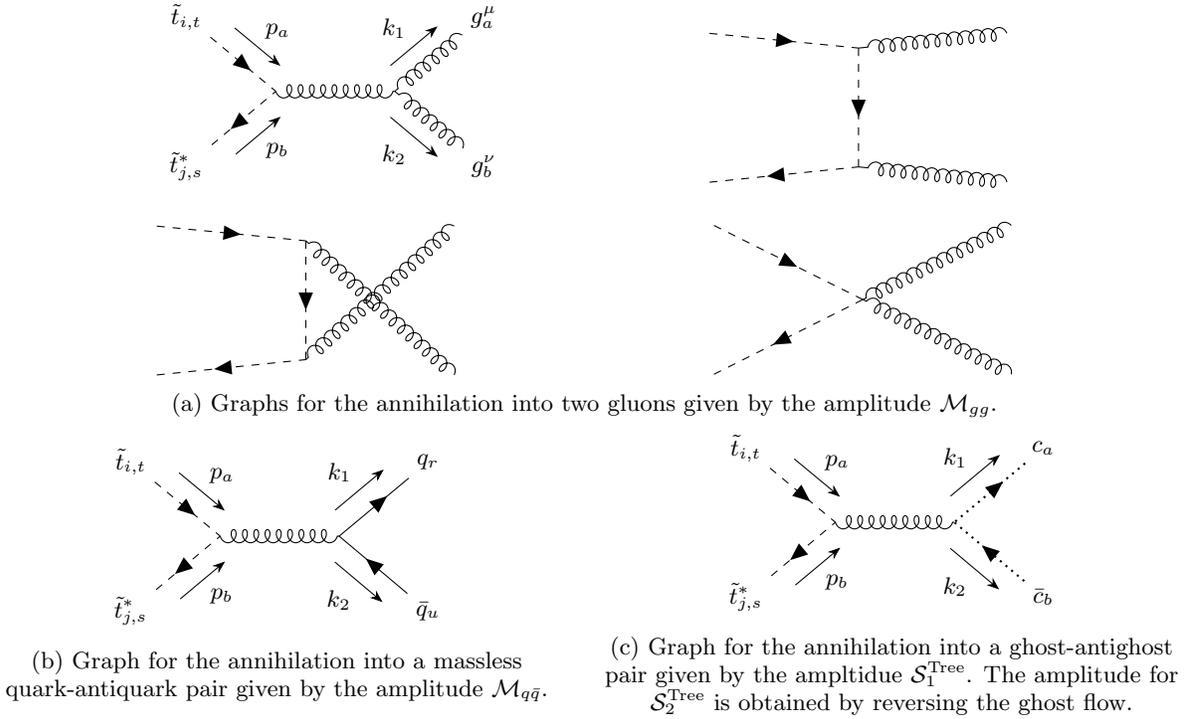

The Feynman diagrams for the leading order process are displayed in \cref{fig:TreeLevelDiags} along with the naming convention for momenta and other relevant indices. 
An important aspect of the processes we investigate is that both initial and final state particles are charged under $\operatorname{SU}(3)_C$. In order to be able to distinguish between attractive and repulsive color potentials in the context of the Coulomb corrections, it is necessary to decompose the tensor product representations under which the two incoming and outgoing particles transform into their respective irreducible representations. The (s)quark-anti(s)quark system  can be decomposed into a color octet and a color singlet
\begin{equation}
 \mathbf{3}\otimes\mathbf{\overline{3}} 	 =\mathbf{8} \oplus \mathbf{1}
 \label{eq:decomposition_inital}
\end{equation}
whereas the decomposition of the two-gluon system reads
\begin{equation}
\mathbf{8}\otimes\mathbf{8}=\mathbf{1}\oplus\mathbf{8}_S \oplus\mathbf{8}_A \oplus\mathbf{\overline{10}}\oplus\mathbf{10}\oplus\mathbf{27}.
\end{equation}
For the decomposition of the tree-level scattering amplitudes
\begin{subequations}
\begin{align}
    \mathcal{M}_{gg}^{\text{Tree}} &= \sum_{\mathbf{R}} c^{[\mathbf{R}]}_{gg} \mathcal{M}^{\text{Tree}}_{ gg,[\mathbf{R}]} \\
    \mathcal{M}_{q\bar{q}}^{\text{Tree}} &= \sum_{\mathbf{R}} c^{[\mathbf{R}]}_{q\bar{q}} \mathcal{M}^{\text{Tree}}_{ q\bar{q},[\mathbf{R}]}
\end{align}
\end{subequations}
into equivalent irreducible representations $\mathbf{R}$ that appear simultaneously in the initial as well as final state, the orthogonal and normalized multiplet basis elements $c^{[\mathbf{R}]}$ spanning the invariant subspaces $\mathbf{R}$ from Ref. \cite{Beneke:2009rj} can be used:
\begin{subequations}
\begin{align}
	c_{gg}^{[\mathbf{1}]}&= \frac{1}{\sqrt{N_c (N_c^2 -1)}}\delta_{st} \delta_{ab}\\
	c_{gg}^{[\mathbf{8_S}]}&= \sqrt{\frac{2 N_c}{C_F(N_c^2-4)}} d_{abc} T^c_{st}\\
	c_{gg}^{[\mathbf{8_A}]}&= i \sqrt{\frac{1}{N^2_c C_F}} f_{abc} T^c_{st} 
\end{align}
\end{subequations}
as well as
\begin{subequations}
\begin{align}
    &c^{[\mathbf{1}]}_{q\bar{q}} = \frac{1}{N_c} \delta_{st} \delta_{ur} \\
     &c^{[\mathbf{8}]}_{q\bar{q}} = \frac{1}{\sqrt{N_c^2-1}} \left(\delta_{su} \delta_{tr}-\frac{1}{N_c} \delta_{st} \delta_{ur}\right)
\end{align}
\end{subequations}
with $C_F=\nicefrac{(N_c^2-1)}{2 N_c}$ and $N_c=3$. 

Another important aspect in a non-Abelian theory is the
treatment of internal and external polarization states. In order to include only the physical external gluon states in the transition probability, we consider two different computational approaches where we use the Feynman gauge for internal gluon lines within both possibilities. The first one is to explicitly sum only the transverse polarizations with the help of the completeness relation
\begin{multline}
   \sum_{T} \epsilon^{\mu\ast}_T(k)  \epsilon^\nu_T (k)  = -g^{\mu\nu}+\frac{k^\mu n^\nu+k^\nu n^\mu}{n \cdot k} - n^2 \frac{k^\mu k^\nu}{(n\cdot k)^2} \label{eq:polSum}
\end{multline}
which holds as an algebraic relation independently of the gauge fixing condition used for the internal propagators and where $n$ is an arbitrary direction in momentum space that fulfills $n\cdot k\neq 0$ and $\epsilon_T(k)\cdot n=0$. For some $n$ with $n^2=0$ this is also referred to as the light-cone gauge. As there appear only two external gluons in the tree-level process, it is instructive to choose $n$ as the momentum of the respective other gluon. The second possibility is to use $-g^{\mu\nu}$ as polarization sum and subtract the longitudinal polarizations by using ghosts. To arrive at the corresponding expression, we derive the two Slavnov-Taylor identities
\begin{subequations}
\label{eq:STI-Tree}
\begin{align}
k_{1}^\mu \mathcal{M}_{gg,\mu\nu}^{\text{Tree}}&=-k_{2,\nu} \mathcal{S}_{1}^{\text{Tree}} \label{eq:ghost1_S}\\
k_{2}^\nu \mathcal{M}_{gg,\mu\nu}^{\text{Tree}}&=-k_{1,\mu} \mathcal{S}_{2}^{\text{Tree}} \label{eq:ghost2_S}
\end{align}
\end{subequations}
from the invariance of a general $n$-point function in SUSY-QCD under Becchi-Rouet-Stora (BRS) transformations \cite{BECCHI1976287,Kugo:1977yx}. Consequently, \cref{eq:STI-Tree} allows to replace the longitudinal polarizations corresponding to all the terms proportional to $k_1$ and $k_2$ in \cref{eq:polSum} with ghost amplitudes. This gives for the squared matrix element summed over final-state polarizations
\begin{equation}
    \mathcal{M}_{gg,\mu\nu}^{\text{Tree}}   (\mathcal{M}_{gg}^{\text{Tree}\ast})^{\mu\nu}-| \mathcal{S}_1^{\text{Tree}}|^2-| \mathcal{S}_2^{\text{Tree}}|^2.
\end{equation}
The fermion spin sum for the quark-antiquark final state is performed in the usual way.
After averaging (summing) over initial- (final-) state colors and performing the remaining phase-space integration, we obtain for the color-decomposed tree-level cross sections describing the annihilation into two gluons the expressions
\begin{align*}
(\sigma v)^{\text{Tree}}_{ gg,[\mathbf{1}]}&=\frac{16 \pi \alpha_s^2}{27 s \beta}\left[\beta(1+\rho)+\rho(\rho-2)\arctanh(\beta)\right] \\
(\sigma v)^{\text{Tree}}_{gg,[\mathbf{8}_S]}&=\frac{5}{2} (\sigma v)^{\text{Tree}}_{gg,[\mathbf{1}]} \\
(\sigma v)^{\text{Tree}}_{gg,[\mathbf{8}_A]}&=\frac{8 \pi \alpha_s^2}{9 s \beta}\left[\beta(1+8 \rho)-3\rho(\rho+2)\arctanh(\beta)\right]
\end{align*}
with $\rho=\nicefrac{4 m^2_{\tilde{q}}}{s}$ and $\beta=\sqrt{1-\rho}$ where $v=2\beta$ corresponds to the relative velocity of the incoming squark-antisquark pair in the c.m. system and $s=(p_1+p_2)^2$ to the squared c.m. energy. 
Only one color channel contributes to the annihilation into a massless quark-antiquark pair giving the cross section
\begin{equation}
(\sigma v)^{\text{Tree}}_{q\bar{q},[\mathbf{8}]} =\frac{16\pi \alpha_s^2\beta^2}{27 s}.
\end{equation}
As we have to combine both processes at NLO, we define already at tree-level
\begin{equation}
    (\sigma v)^{\text{Tree}}=(\sigma v)^{\text{Tree}}_{gg}+N_f (\sigma v)^{\text{Tree}}_{q\bar{q}}
\end{equation}
where $N_f=4$ corresponds to the number of effectively massless quark flavors.

\subsection{\label{sec:scenario}Reference scenario and numerical discussion}
To illustrate the importance of stop annihilation into gluons, we introduce the reference scenario given in \cref{tab:scenario} which has been found by performing a random scan in the MSSM with 19 free parameters considering the most important experimental constraints from searches for supersymmetry. 
\renewcommand{\arraystretch}{1.4}
\begin{table*}
\centering
\caption{\label{tab:table3}$\overline{\text{DR}}$ parameters for the reference scenario in the pMSSM-19 defined at the scale $Q_{\text{SUSY}}=\sqrt{m_{\tilde{t}_1} m_{\tilde{t}_2}}$ where $m_{\tilde{t}_1}$ and $m_{\tilde{t}_2}$ are in this case the $\overline{\text{DR}}$ tree-level masses, the associated pole masses of relevant particles, the bino contribution $Z_{11}$ to $\tilde{\chi}^0_1$ and the neutralino relic density. All dimensionful quantities are given in $\si{\giga\electronvolt}$.}
\begin{ruledtabular}
\begin{tabular}{cccccccccc}
$M_1$ & $M_2$ & $M_3$ & $M_{\tilde{l}_L}$ & $M_{\tilde{\tau}_L}$ & $M_{\tilde{l}_R}$ & $M_{\tilde{\tau}_R}$ & $M_{\tilde{q}_L}$ & $M_{\tilde{q}_{3L}}$ & $M_{\tilde{u}_R}$ \\ 
1437.9 & 2739.6 & 3079.5 & 4034.1 & 3620.2 & 4075.12 & 2605.9 &  1773.2 & 2172.7 & 1816.1 \\ \hline
$M_{\tilde{t}_R}$ & $M_{\tilde{d}_R}$ & $M_{\tilde{b}_R}$ & $A_t$ & $A_b$ & $A_\tau$ & $\mu$ & $m_{A^0}$ & $\tan\beta$ & $Q_{\text{SUSY}}$ \\ 
1424.3 & 1926.8 & 2913.0 & 2965.3 & 3050.7 & 2880.3 & -1880.8 & 3742.2 & 34.9 & 1756.4
\end{tabular}
\vspace{2mm}
\begin{tabular}{ccccccccccc}
$m_{\tilde{\chi}^0_1}$ & $m_{\tilde{\chi}^0_2}$ & $m_{\tilde{\chi}^{\pm}_1}$ & $m_{\tilde{t}_1}$ & $m_{\tilde{t}_2}$ & $m_{\tilde{g}}$ & $m_{\tilde{\tau}_1}$ & $m_{h^0}$ & $m_{H^0}$ & $Z_{11}$ & $\Omega_{\tilde{\chi}^0_1}h^2$ \\ 
1435.7 & 1884.4 & 1882.9 & 1446.3 & 2248.0 & 3059.3 & 2613.5 & 124.0 & 3742.9 & 0.9976 &0.1201
\end{tabular}  
\end{ruledtabular}
\label{tab:scenario}
\end{table*}
\renewcommand{\arraystretch}{1.0}
For this scan and throughout our analysis \texttt{SoftSUSY 4.1.9} \cite{Allanach:2001kg,Allanach:2014nba,Allanach:2016rxd,Allanach:2017hcf} is used for the generation of the mass spectrum and mixing parameters with the option of including three-loop corrections to the mass of the CP-even Higgs boson $h^0$ provided by \texttt{Himalaya 1.0} \cite{Kant:2010tf,Harlander:2017kuc} turned on. Only those points that obey the Higgs mass limit $\SI{123}{ \giga\electronvolt}<m_{h^0}<\SI{127}{\giga\electronvolt}$, feature the neutralino as lightest supersymmetric particle (LSP) and a stop as next-to-lightest supersymmetric particle (NLSP) are taken into account. We use \texttt{SModelS 2.2.0} \cite{Kraml:2013mwa,Ambrogi:2017neo,Ambrogi:2018ujg,Heisig:2018kfq,Dutta:2018ioj} and \texttt{SUSY-AI} \cite{Caron:2016hib} to exclude points that have been ruled out by LHC searches for supersymmetry. The consistency of the Higgs sector with measurements from LEP, Tevatron and the LHC is additionally checked with \texttt{HiggsBounds 5.5.0} \cite{Bechtle:2020pkv} and \texttt{HiggsSignals 2.3.0} \cite{Bechtle:2013xfa}. The module in \texttt{MicrOMEGAs-5.2.13} \cite{Barducci:2016pcb} is used to check against constraints from dark matter direct detection experiments. However, unless stated otherwise we use throughout this paper \texttt{MicrOMEGAs 2.4.1} \cite{Belanger:2001fz,Belanger:2006is} with the standard \texttt{CalcHEP} implementation of the MSSM for the computation of the relic density and the contributions of different (co)annihilation channels.

The latter are shown in \cref{tab:channels} for the chosen reference scenario.
\begin{figure*}
    \includegraphics[width=0.45\textwidth]{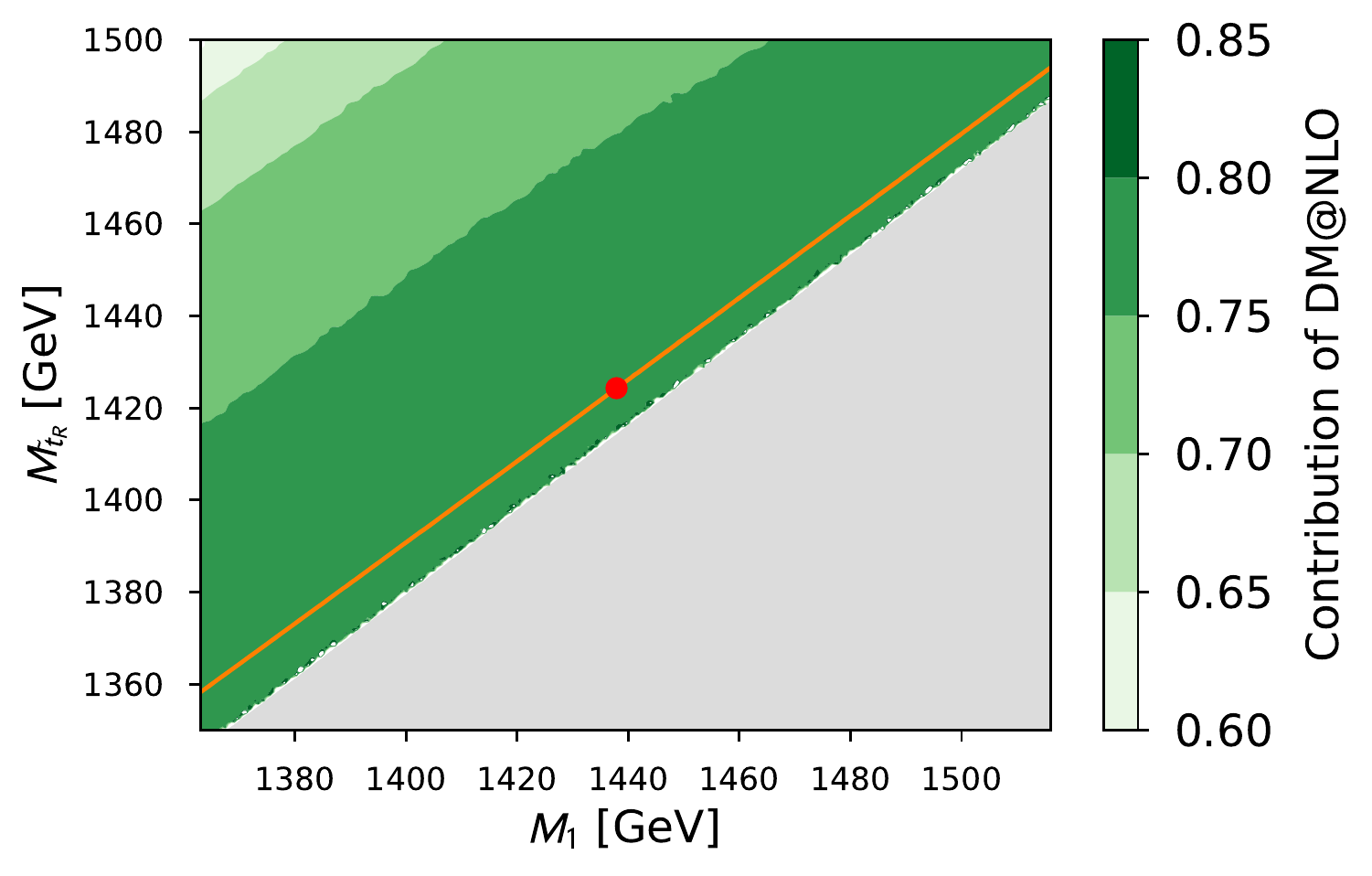}
    \includegraphics[width=0.45\textwidth]{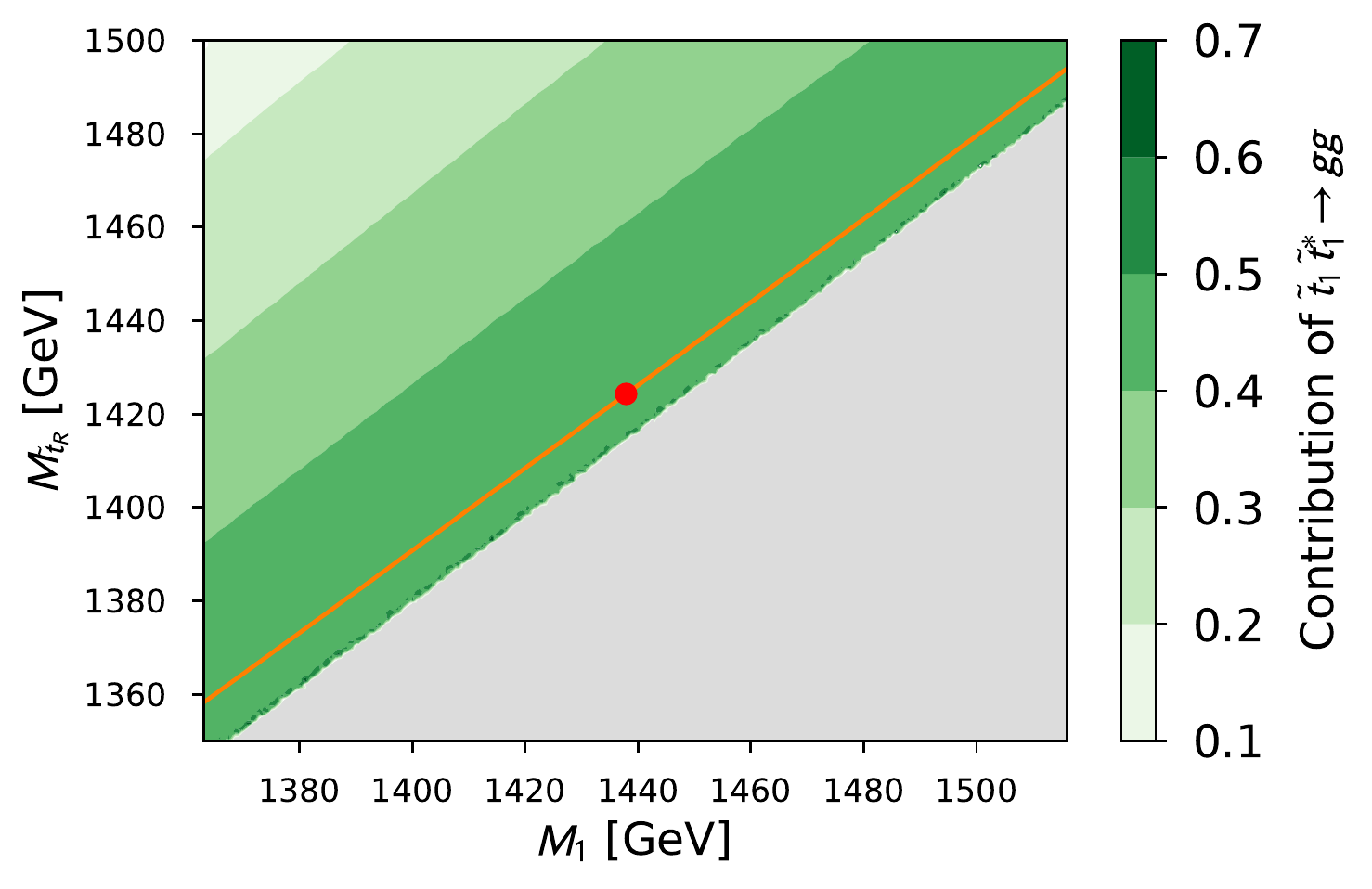}
    \includegraphics[width=0.45\textwidth]{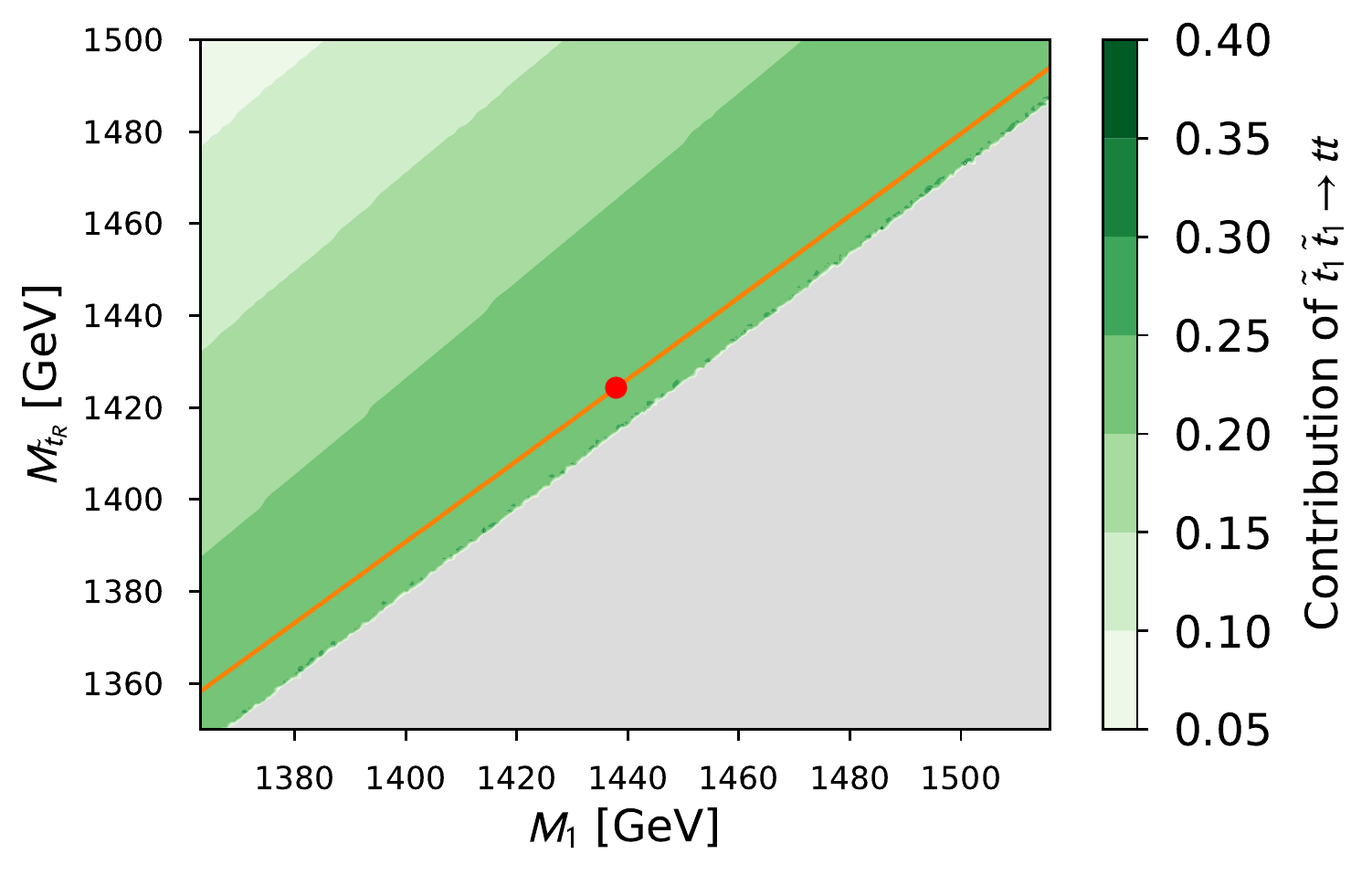}
    \includegraphics[width=0.45\textwidth]{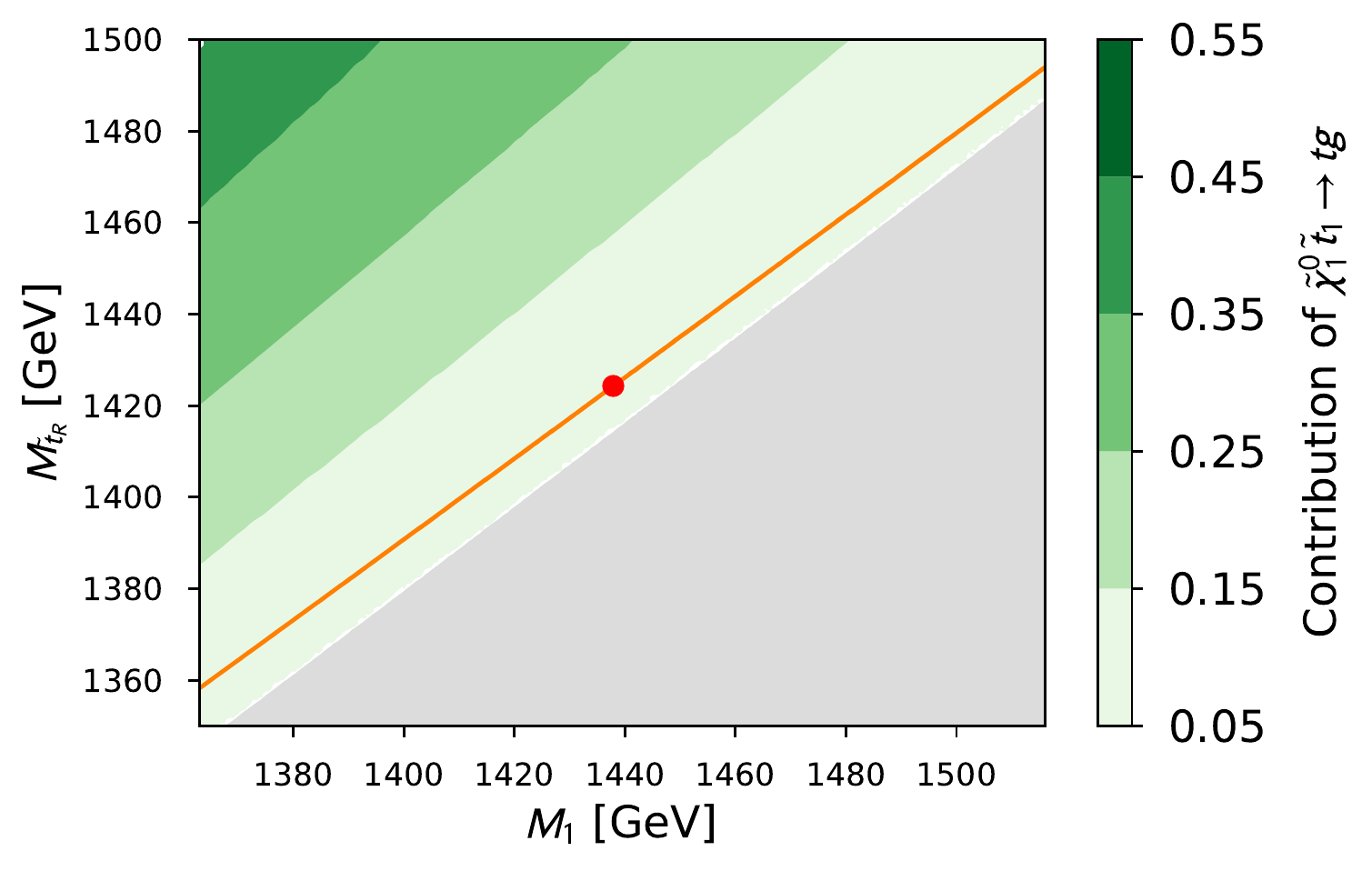}
    \caption{Contribution of relevant processes that can be corrected by \texttt{DM@NLO} to the effective annihilation cross section in the $M_1$-$M_{\tilde{t}_R}$ plane around the chosen reference scenario which is highlighted with a red dot. The region where the neutralino is not the LSP is marked in gray. The orange band indicates the parameter region that is consistent with the \emph{Planck} measurement given in \cref{eq:omh2} at the $2\sigma$ confidence level based on the tree-level cross sections provided by \texttt{CalcHEP}.}
    \label{fig:m1-mtR-MO}
\end{figure*}
\renewcommand{\arraystretch}{1.4}
\begin{table}[b]
\caption{\label{tab:channels}
Dominant annihilation channels contributing to $\left(\Omega h^2\right)^{-1}$ for the scenario in \cref{tab:scenario}. Further contributions below $\SI{2}{\percent}$ are omitted.}
\begin{ruledtabular}
\begin{tabular}{cc}
Channel & Contribution \\ \hline
$\tilde{t}_1 \ \tilde{t}_1^\ast \rightarrow  g \ g$ & $\SI{47}{\percent}$   \\ 
$\tilde{t}_1 \ \tilde{t}_1 \rightarrow  t \ t$  &   $\SI{23}{\percent}$  \\
$\tilde{\chi}^0_1 \ \tilde{t}_1\rightarrow g \ t $ & $\SI{7}{\percent}$  \\
$\tilde{t}_1 \ \tilde{t}_1^\ast \rightarrow \gamma \ g$ & $\SI{7}{\percent}$  \\
$\tilde{t}_1 \ \tilde{t}_1^\ast \rightarrow t \ \bar{t}$ &  $\SI{5}{\percent}$  \\
$\tilde{t}_1  \ \tilde{t}_1^\ast \rightarrow Z^0 \ g $ & $\SI{2}{\percent}$ \\ \hline
\texttt{DM@NLO} total \cite{Schmiemann:2019czm,Harz:2014tma} & $\SI{77}{\percent}$
\end{tabular}
\end{ruledtabular}
\end{table}
\renewcommand{\arraystretch}{1.0}
The largest contribution comes with $\SI{47}{\percent}$ from stop-antistop annihilation into gluons followed in decreasing order by stop pair-annihilation into top quarks and neutralino-stop coannihilation into a gluon and a top quark which have been previously analyzed in \cite{Schmiemann:2019czm} and \cite{Harz:2014tma}, respectively. In total, \texttt{DM@NLO} provides full one-loop SUSY-QCD corrections to $\SI{77}{\percent}$ of the effective cross section in \cref{eq:eff_XSec}.

The scenario features a bino-like neutralino which is not surprising as large wino and higgsino components would lead to other gauginos being the NLSP and the mass difference between the neutralino and the lightest stop is approximately $\SI{11}{\giga\electronvolt}$. The gluino and slepton sector are chosen to be much heavier than the stop sector to ensure that they do not influence the phenomenology discussed here. 
In \Cref{fig:m1-mtR-MO}, the relative contributions of the three most important channels to the relic density are displayed in the $M_1$-$M_{\tilde{t}_R}$ mass plane in different shades of green. We choose these two parameters as the lightest neutralino is mostly bino-like and its mass is consequently predominately given by $M_1$. The $M_{\tilde{t}_R}$ parameter enters the tree-level expression of the $\tilde{t}_1$ mass and therefore these two parameters correspond to a scan in the $m_{\tilde{\chi}^0_1}$-$m_{\tilde{t}_1}$ mass plane which in turn allows to investigate the dependence of the relic density on the LSP-NSLP mass difference.  For larger mass splittings between the lightest neutralino and the stop coannihilation becomes the dominant channel whereas for small mass splittings annihilation of stops is the dominant contribution. In addition, the region where the neutralino accounts for the whole dark matter content in the universe and lies within the $2\sigma$ range of the experimental value is marked in orange. This region follows an almost straight line parallel to the boundary where the neutralino is no longer the LSP.

\begin{figure*}
\includegraphics[width=0.6\textwidth]{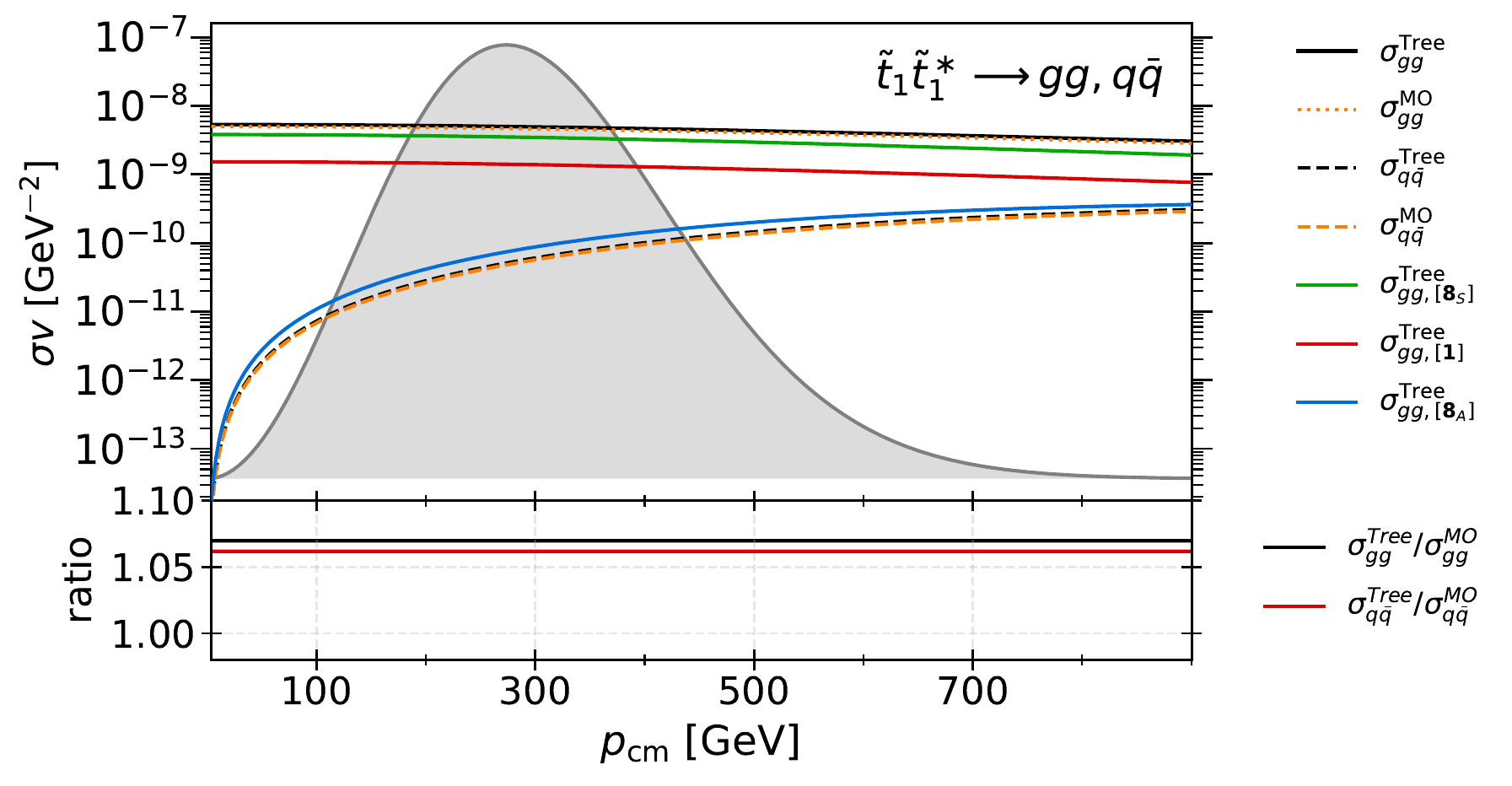} 
\caption{Leading order cross sections times velocity introduced in \cref{sec:LO-XSec} as provided by $\texttt{DM@NLO}$ as well the corresponding results from \texttt{CalcHEP} indicated with the superscript MO. All cross sections are displayed in dependence of the CM momentum for the chosen reference scenario.}
\label{fig:LO-XSec}
\end{figure*}

With the knowledge that stop annihilation into gluons is important for large regions around the reference scenario, we turn now to the numerical comparison between our leading order cross sections for the two processes in \cref{eq:processes} and the ones from \texttt{MicrOMEGAs 2.4.1} which are all shown in \cref{fig:LO-XSec}. As a reminder that the values of of the cross section impacts the relic density only in a limited energy range, the Boltzmann distribution which is involved in the computation of the thermally averaged cross section at freeze-out temperature is shown in gray in arbitrary units. One observes that our result is about $\SI{6}{\percent}$ larger for both processes which has two reasons. Firstly, we set the renormalization scale which enters at tree-level only through the strong coupling to $\mu_R=Q_{\text{SUSY}}$ whereas \texttt{MicrOMEGAs 2.4.1} sets the scale to twice the dark matter mass $\mu_{\text{MO}}=2 m_{\tilde{\chi}^0_1}$ which is larger than $\mu_R$ for the investigated scenario and therefore corresponds to a smaller strong coupling. {Our choice for $\mu_R$ is motivated by the fact that the besides the masses of the virtual particles in the loop, the process contains only two  important scales: the mass of the lightest stop and the collisional energy $s$. Since most annihilations take place between $s = 4 m^2_{\tilde{t}_1}$ and the peak of the velocity distribution at $s \sim (\SI{3}{\tera\electronvolt})^2$, $Q_{\text{SUSY}}$ is a suitable choice for the renormalization scale to avoid large logarithms.} Secondly, \texttt{MicrOMEGAs 2.4.1} calculates the running of $\alpha_s$ in the $\overline{\text{MS}}$-scheme using the three-loop formula in Ref. \cite{ParticleDataGroup:2004fcd} with six active flavors and the SM particle content only whereas \texttt{DM@NLO} uses the four-loop formula from Ref. \cite{Vermaseren:1997fq} in the $\overline{\text{DR}}$-scheme \cite{Harlander:2005wm} with six active flavors and contributions from the complete MSSM mass spectrum \cite{Bauer:2008bj}. Considering only these two differences in the computation, the ratio should be identical for both processes, but this is not the case as \texttt{MicrOMEGAs} also takes into account electroweak contributions with a photon or a $Z^0$ propagator for the process with a quark-antiquark pair in the final state. {The corresponding electroweak diagrams are not included in our calculation since the process with massless quarks is numerically insignificant for the relic density as well as the tree-level cross section compared to the annihilation into gluons as visible in \cref{tab:channels} and \cref{fig:LO-XSec} and was only added for consistency to achieve an infrared finite result.}

Through comparison of the different color contributions to the combined leading order cross section depicted in \cref{fig:LO-XSec} with the partial wave expansion 
\begin{equation}
    \sigma v  =  s_0 + v^2 s_1 + \mathcal{O}(v^4) 
\end{equation}
of a general velocity-weighted annihilation cross section $\sigma v$, it becomes apparent that the singlet and symmetric octet contributions to the cross section with two external gluons are dominated by the $S$-wave component $s_0$ since they remain almost constant in $v$, whereas the antisymmetric octet part of the same process and the octet contribution to the quark-antiquark process take an inferior role and are suppressed at threshold corresponding to the $S$-wave and $P$-wave component $s_1$.

\section{\label{sec:compDetails}Computational details of the radiative corrections}
In this section, we discuss the technical details of our SUSY-QCD corrections at $\order{\alpha_s}$ as well as the Sommerfeld enhancement. The NLO cross section
\begin{equation}
    (\sigma v)^{\text{NLO}} = (\sigma v)^{\text{Tree}} + \Delta(\sigma v)^{\text{NLO}}
\end{equation}
with the NLO correction 
\begin{equation}
   \Delta\sigma^{\text{NLO}} = \int_2 \dd\sigma^{\text{V}}+\int_3\dd\sigma^{\text{R}} 
\end{equation}
consists of virtual $\dd\sigma^{\text{V}}$ and real corrections $\dd\sigma^{\text{R}}$. The integration domain of the integrals refers to the number of final-state particles.
Both contributions have been calculated and verified with the publicly available tools \texttt{FeynArts 3} \cite{Hahn:2000kx}, \texttt{FeynCalc 9} \cite{Shtabovenko_2020}, \texttt{Tracer} \cite{Jamin:1991dp} and \texttt{FormCalc 9} \cite{Hahn:2000jm}. 

\subsection{\label{sec:virtual}Virtual corrections and renormalization}
\begin{figure*}[ht]
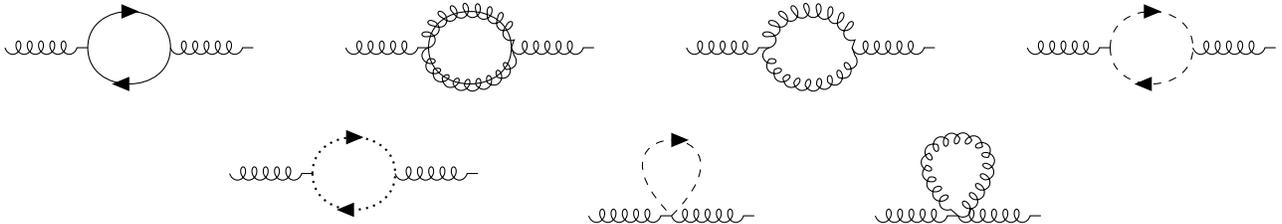

    \centering
 
\caption{One-loop contributions to the gluon self-energy.}
\label{fig:gluon-self-energies}
\end{figure*}

\begin{figure*}[ht]
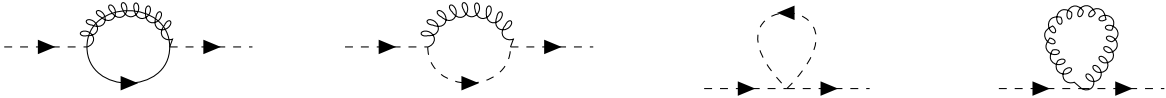

    %
\caption{Contributions to the squark self-energy at one-loop.}
\label{fig:squark-self-energies}
\end{figure*}

\begin{figure*}[ht]
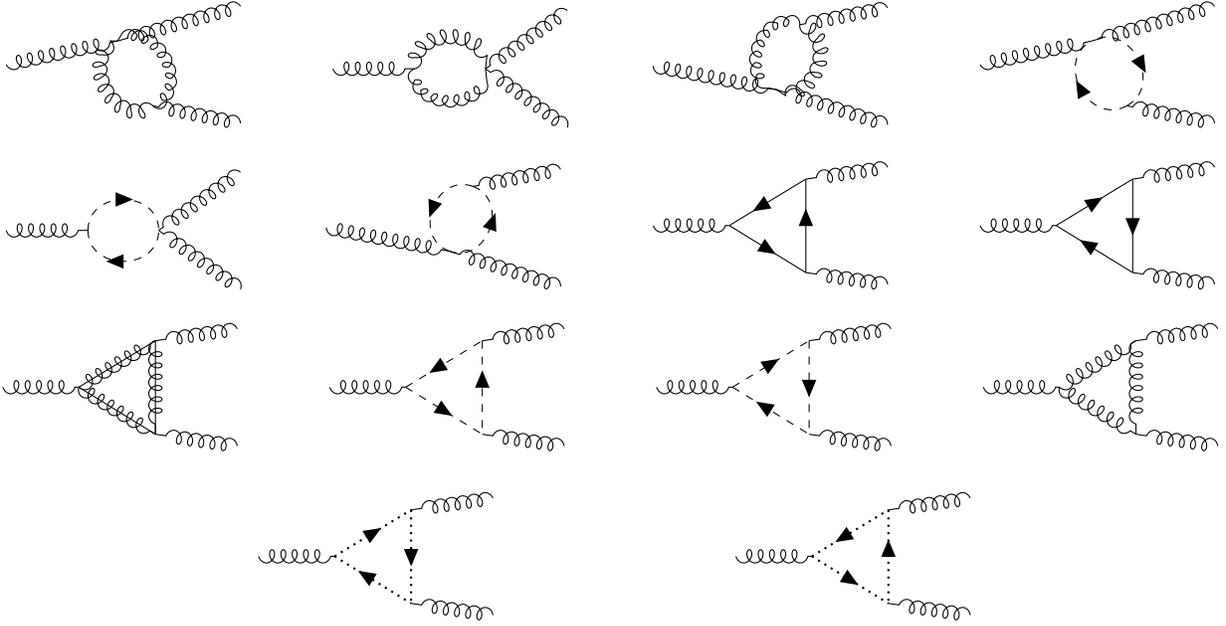

    \centering
%
 
\caption{One-loop contributions to the triple-gluon vertex.}
\label{fig:triple-gluon-vertex}
\end{figure*}
\begin{figure*}[ht]
\centering
%

\caption{One-loop contributions to the squark-gluon vertex.}
\label{fig:diags_squark-gluon-vertex}
\end{figure*}
\begin{figure*}[ht]
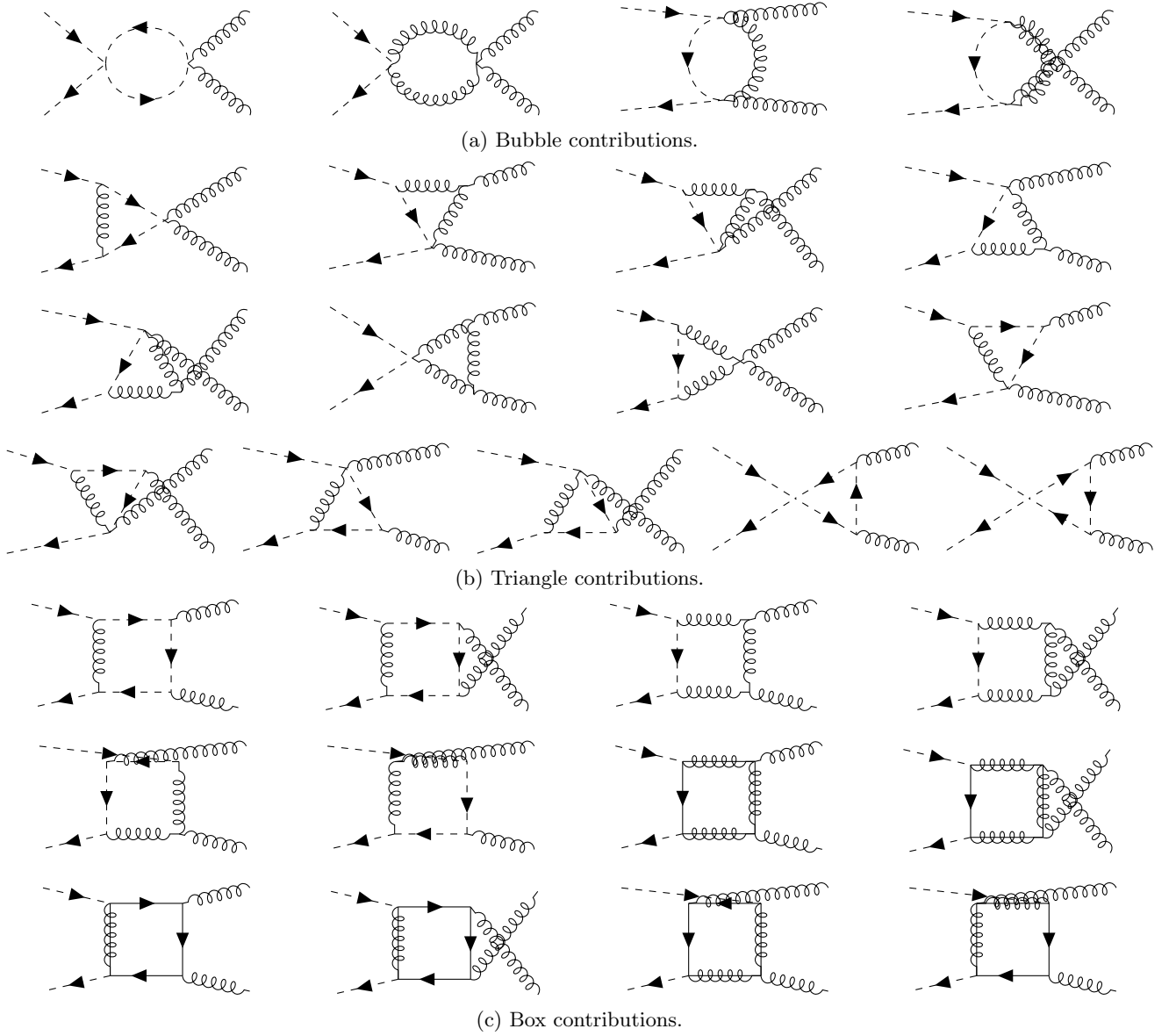

\centering
\begin{subfigure}[c]{0.99\textwidth}	
%
 
\caption{Bubble contributions.}
\label{fig:four-vertex-Bubbles}
\end{subfigure}

\begin{subfigure}[c]{0.99\textwidth}	
%
\subcaption{Triangle contributions.}
\label{fig:four-verteX-triangles}
\end{subfigure}

 \begin{subfigure}[c]{0.99\textwidth}	
 \centering
%
\subcaption{Box contributions.}
\label{fig:four-gluon-squark-all-diags-boxes}
\end{subfigure}
\caption{One-loop corrections to the four-gluon-squark vertex.}
\label{fig:four-vertex}
\end{figure*}
\begin{figure*}[ht]
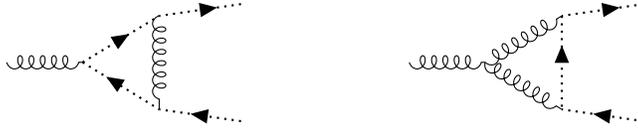

\centering
%
\caption{One-loop contributions to the ghost-gluon vertex.}
\label{fig:ghost-vertex-diags}
\end{figure*}
\begin{figure*}[ht]
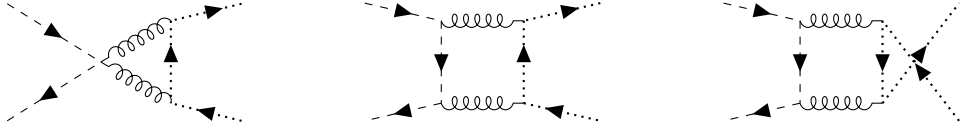
 
\centering
%
\caption{Triangle and box corrections to the ghost process $\mathcal{S}_1^{\text{NLO}}$ which do not have a tree level analogue. The diagrams for $\mathcal{S}_2^{\text{NLO}}$ can be obtained by reversing the ghost flow}
\label{fig:Box_corrections_ghost}
\end{figure*}
\begin{figure*}[ht]
\centering
%

\caption{One-loop contributions to the quark-gluon vertex.}
\label{fig:diags_quark-gluon}
\end{figure*}

\begin{figure*}[ht]
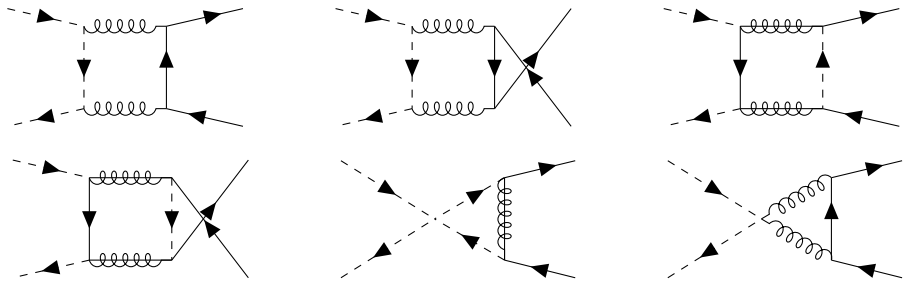

 \centering	
 %
 
\caption{Box and triangle diagrams associated with stop-antistop annihilation into light quarks.}
\label{fig:quark-boxes}
\end{figure*}

The virtual amplitudes consist of propagator (self-energy), vertex and box corrections. Naively one might assume that the box corrections for the process with two final-state gluons are independent and UV finite on their own. However, they turn out to be UV divergent and fall under the renormalization of the four-squark-gluon vertex. All corresponding Feynman diagrams are shown in \cref{fig:gluon-self-energies,fig:squark-self-energies,fig:triple-gluon-vertex,fig:diags_squark-gluon-vertex,fig:four-vertex,fig:ghost-vertex-diags,fig:Box_corrections_ghost,fig:diags_quark-gluon,fig:quark-boxes}. We subtract the longitudinal gluon polarizations again through ghosts, i.e. the interference of the tree-level matrix element with the virtual amplitudes for the process with two gluons in the final state summed over the final-state polarizations can be written as
\begin{align}
2  \Re\left[(\mathcal{M}_{gg}^{\text{Tree}\ast})^{\mu\nu} \mathcal{M}_{gg,\mu\nu}^{\text{NLO}} -\mathcal{S}_1^{\text{Tree}\ast}
\mathcal{S}_1^{\text{NLO}}- \mathcal{S}_2^{\text{Tree}\ast} \mathcal{S}_2^{\text{NLO}} \right]
\end{align}
where some of the ghost corrections making up the ghost amplitudes $\mathcal{S}_i^{\text{NLO}}$ ($i=1,2$) are shown in \cref{fig:ghost-vertex-diags,fig:Box_corrections_ghost}.
These diagrams are regulated dimensionally in $D=4-2\varepsilon$ dimensions within the supersymmetry preserving four-dimensional helicity scheme \cite{Siegel:1979wq,Stockinger:2005gx,Signer:2008va} so that UV and IR divergences appear as poles of the form $\varepsilon^{-1}$ and $\varepsilon^{-2}$. The standard Passarino-Veltman reduction \cite{Passarino:1978jh, Denner:1991kt} is used to express the one-loop amplitudes in terms of the well-known scalar integrals $A_0$, $B_0$, $C_0$, $D_0$ \cite{Dittmaier:2003bc,Ellis:2007qk,Denner:2010tr}. The $\gamma^5$-matrix which enters through the squark-quark-gluino coupling is treated in the naive scheme, i.e. we assume that $\gamma^5$ still anti-commutes with all $\gamma$-matrices in $D$ dimensions. The Levi-Civita symbols that occur then through traces of $\gamma^5$ with four or more $\gamma$-matrices during the evaluation of diagrams with top quarks as virtual particles are directly set to zero since they vanish anyway when being contracted with the external momenta.
The UV divergences that appear in the virtual corrections are removed through the renormalization of fields, masses and the strong coupling. Within our calculation, a hybrid on-shell/$\overline{\text{DR}}$ renormalization scheme is employed where $A_t$, $A_b$, $m_{\tilde{t}_1}$, $m_{\tilde{b}_1}$, $m_{\tilde{b}_2}$ along with the heavy quark masses $m_t$, $m_b$ are treated as independent input parameters so that the mixing angles $\theta_{\tilde{t}_1}$, $\theta_{\tilde{t}_2}$ and the mass of the heavier stop $m_{\tilde{t}_2}$ depend on their definition. The trilinear couplings of the third generation, the bottom quark mass and the strong coupling are renormalized in the $\overline{\text{DR}}$ scheme while the on-shell scheme has been chosen for the top mass and the input squark masses. This particular scheme resembles the RS2 scheme introduced in Ref. \cite{Heinemeyer:2010mm} and was found to be robust over large regions of the parameter space for (co)annihilations involving stops in a series of previous analyses \cite{Harz:2012fz,Harz:2014tma}. Since the renormalization of the gluon and the squark sector as well as the treatment of the bottom mass and the strong coupling have already been discussed in detail in the context of other processes \cite{Herrmann:2014kma,Harz:2014tma,Harz:2012fz}, we will only cover aspects which are new to this calculation in the following such as the renormalization of ghosts and massless quarks.
\subsubsection{Ghost wave-function renormalization}
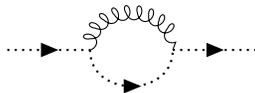
\begin{figure*}[ht]
\centering
\begin{tikzpicture}
    \begin{feynman}
    \vertex (a);
    \vertex[right=1.1 of a] (b);
    \vertex[right=1.1 of b] (c);
    \vertex[right=1.1 of c] (d);
    \diagram*{(a)--[ghost] (b),
    (b) -- [gluon, half left] (c), 
    (b) -- [ghost, half right] (c),
    (c) -- [ghost] (d),
    };
    \end{feynman}
\end{tikzpicture}
\caption{One-loop contribution to the ghost self-energy.}
\label{fig:ghost_self_energy}
\end{figure*}
As ghost and anti-ghost share the same self-energy they can be renormalized with the same wave function renormalization constant $Z_c$. The renormalized fields are then defined as
\begin{align}
&\overline{c}^0_a = \sqrt{Z_c} \overline{c}^R_a \\
&c_a^0 = \sqrt{Z_c} c_a^R
\end{align}
where we need $\delta Z_c$ only up to $\order{\alpha_s}$ which leads to the expansion
\begin{equation}
Z_c = 1 + \delta Z_c.
\end{equation}
 Since the gluon is renormalized in the on-shell scheme, the same scheme is chosen for the ghost. That is, the ghost renormalization constant is obtained by requiring that the ghost Green's function has a unit residue even up to the one-loop level
\begin{equation}
\delta Z_c=-\Re\left.\dot{\Pi}_c\left(p^2\right)\right|_{p^2=0}
\end{equation}
where 
\begin{equation}
\dot{\Pi}_c(p^2)=-\frac{\alpha_s N_c}{8\pi}\left(B_0(p^2,0,0)-1\right).
\end{equation}
denotes the derivative of the ghost self-energy whose only contribution is depicted in \cref{fig:ghost_self_energy}.
 The constant $\delta Z_c$ contains UV and IR divergent parts which read explicitly
\begin{align}
\delta Z_c^{\text{UV}} &= \frac{\alpha_s N_c}{8\pi \varepsilon_{\text{UV}}} \\
\delta Z_c^{\text{IR}} &= -\frac{\alpha_s N_c}{8\pi \varepsilon_{\text{IR}}}.
\end{align}

\subsubsection{Renormalization of the massless quarks}

\begin{figure*}[ht]
\centering
\begin{tikzpicture}
    \begin{feynman}
    \vertex (a);
    \vertex[right=1.1 of a] (b);
    \vertex[right=1.1 of b] (c);
    \vertex[right=1.1 of c] (d);
    \diagram*{(a)--[fermion] (b),
    (b) -- [gluon, half left] (c), 
    (b) -- [fermion, half right] (c),
    (c) -- [fermion] (d),
    };
    \end{feynman}
\end{tikzpicture}
\hspace{0.7cm}
\begin{tikzpicture}
    \begin{feynman}
    \vertex (a);
    \vertex[right=1.1 of a] (b);
    \vertex[right=1.1 of b] (c);
    \vertex[right=1.1 of c] (d);
    \diagram*{(a)--[fermion] (b),
    (b) -- [charged scalar, half left] (c), 
    (b) -- [plain,gluon, half right] (c),
    (c) -- [fermion] (d),
    };
    \end{feynman}
\end{tikzpicture}
\caption{One-loop contributions to the quark self-energy.}
\label{fig:quark_self_energy}
\end{figure*}
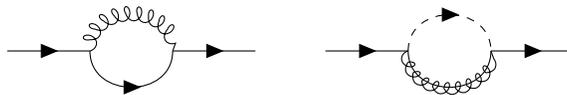
For the renormalization of massless quarks, we introduce the quark wave-function renormalization constants $Z_q^{L/R}$ for each chirality state
\begin{equation}
    q_{L/R} = \sqrt{Z_q^{L/R}} q_{L/R} = (1+\tfrac{1}{2} \delta Z_q^{L/R}) q_{L/R}.
\end{equation}
The renormalization constants are determined in the on-shell scheme which requires the renormalized quark two-point Green's function to have a unit residue. This condition results in the expression 
\begin{equation}
    \delta Z_q^{L/R} = -\Re \Pi^{L/R}_q(0)
\end{equation}
where the function $\Pi^{L/R}_q(q^2)$ appears in the decomposition of the quark self-energy
\begin{multline}
    \Pi_q(p) = \slashed{p}\left[ P_L  \Pi_q^L(p^2) + P_R \Pi_q^R(p^2)\right] \\ + \Pi_q^{S,L}(p^2) P_L+ \Pi_q^{S,R}(p^2) P_R
\end{multline}
whose two contributing Feynman diagrams are shown in \cref{fig:quark_self_energy}. The resulting constants contain the UV and IR divergent parts
\begin{align}
\delta Z_q^{\text{UV}} &= -\frac{\alpha_s C_F}{2\pi \varepsilon_{\text{UV}}} \\
\delta Z_q^{\text{IR}} &= \frac{\alpha_s C_F}{4\pi \varepsilon_{\text{IR}}},
\end{align}
where the superscripts indicating the left/right-handed chirality states are dropped here for simplicity. 

\subsection{\label{sec:real}Real corrections}
\begin{figure*}[ht]
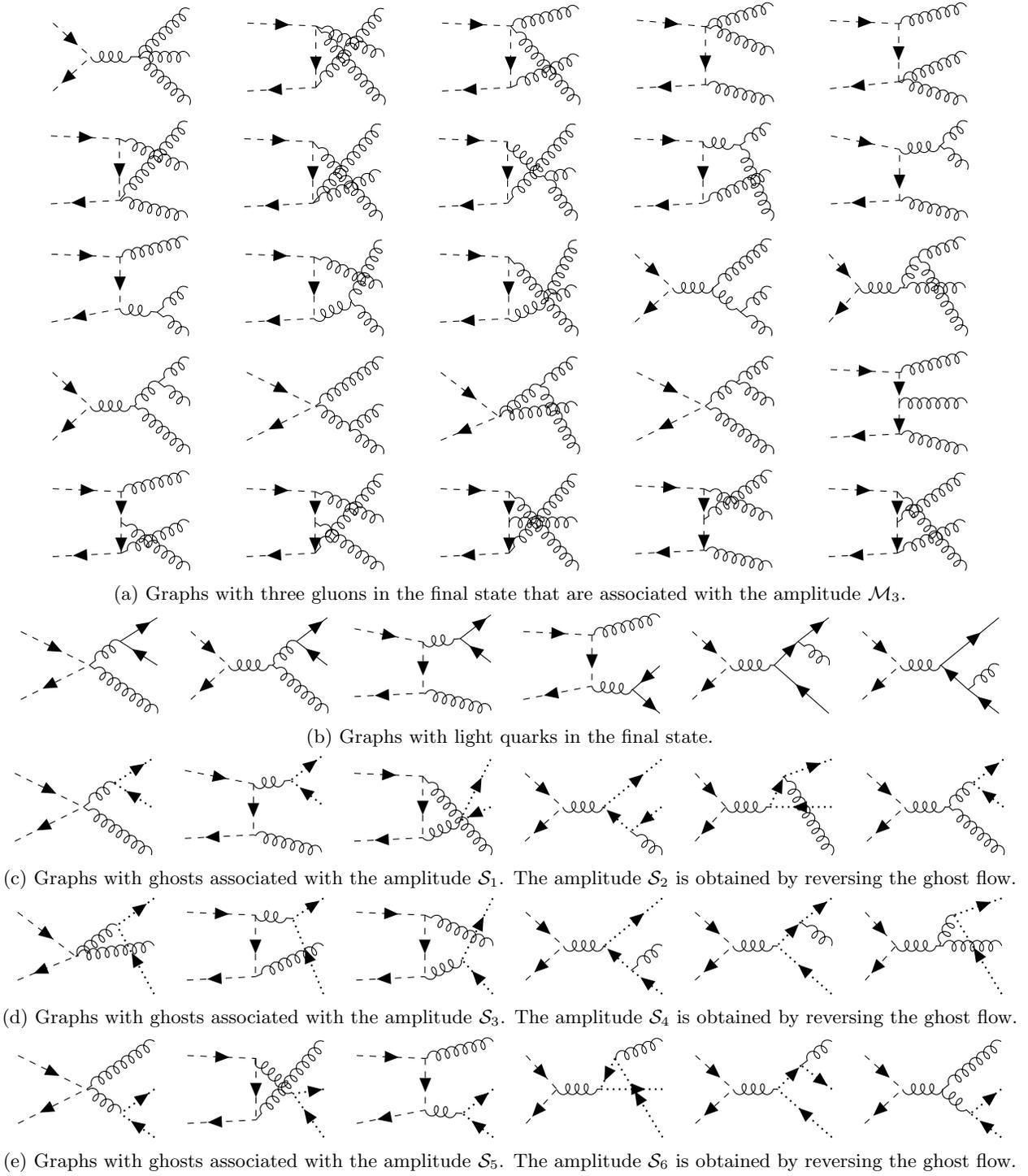

\begin{subfigure}[c]{\textwidth}	
          

\subcaption{Graphs with three gluons in the final state that are associated with the amplitude $\mathcal{M}_3$.}
\label{fig:real_3g}
\end{subfigure}
\begin{subfigure}[c]{\textwidth}	
    \centering 
     %
 
\subcaption{Graphs with light quarks in the final state.}
\label{fig:real_qqbar}
\end{subfigure}
\begin{subfigure}[c]{0.99\textwidth}	
\centering
%
\subcaption{Graphs with ghosts associated with the amplitude $\mathcal{S}_1$. The amplitude $\mathcal{S}_2$ is obtained by reversing the ghost flow.}
\label{fig:ghost12}
\end{subfigure}

    \begin{subfigure}[c]{0.99\textwidth}		
        \centering	
%
\subcaption{Graphs with ghosts associated with the amplitude $\mathcal{S}_3$. The amplitude $\mathcal{S}_4$ is obtained by reversing the ghost flow.}
\label{fig:ghost34}
    \end{subfigure}
    \begin{subfigure}[c]{0.99\textwidth}		
        \centering	
%
  
\subcaption{Graphs with ghosts associated with the amplitude $\mathcal{S}_5$. The amplitude $\mathcal{S}_6$ is obtained by reversing the ghost flow.}
\label{fig:ghost56}
    \end{subfigure}
\caption{Real emission diagrams for stop annihilation into gluons and light quarks.}
\end{figure*}

The infrared divergences in the virtual corrections are compensated by including the real emission processes 
\begin{equation}
\tilde{t}_1 \tilde{t}_1^{\ast}\longrightarrow g_a^\mu\left(k_1\right)+g_b^\nu\left(k_2\right)+g_c^\rho\left(k_3\right) \label{eq:process_3g}
\end{equation}
and
\begin{equation}
 \tilde{t}_1 \tilde{t}_1^{\ast} \longrightarrow q_r\left(k_1\right)+\overline{q}_u\left(k_2\right)+g_a^\mu\left(k_3\right)
 \label{eq:process_gqqbar}
\end{equation}
with $q \in \{u,d,c,s\}$ being an effectively massless quark and where the initial squarks carry the same labels as in \cref{fig:TreeLevelDiags}.
The corresponding Feynman diagrams are shown in \cref{fig:real_3g,fig:real_qqbar} where the momenta of the gluons in the first process have to be read from top to bottom starting with $k_1$.
As in the tree-level calculation, we use $-g^{\mu\nu}$ for the gluon polarization sum and subtract the longitudinal polarizations with ghosts as asymptotic states. In order to arrive at the corresponding expression, we proceed as sketched in \cref{sec:LO-XSec} by deriving the following two sets of Ward identities from BRS-invariance
\begin{subequations}
\label{eq:WardM3}
\begin{align}
k_{1,\mu} \mathcal{M}_3^{\mu\nu\rho} &= -k_2^\nu \mathcal{S}_1^\rho -  k_3^\rho \mathcal{S}_3^\nu   \\
k_{2,\nu} \mathcal{M}_3^{\mu\nu\rho} &= -k_1^\mu \mathcal{S}_2^\rho - k_3^\rho \mathcal{S}_6^\mu \\
k_{3,\rho} \mathcal{M}_3^{\mu\nu\rho} &= -k_1^\mu \mathcal{S}_4^\nu -  k_2^\nu \mathcal{S}_5^\mu
\end{align} 
\end{subequations}
and
\begin{subequations}
\label{eq:WardS}
\begin{align}
k_{2,\nu}\mathcal{S}_4^\nu &= k_{3,\rho} \mathcal{S}_2^\rho  \\
k_{1,\mu}\mathcal{S}_5^\mu &= k_{3,\rho} \mathcal{S}_1^\rho \\
k_{1,\mu}\mathcal{S}_6^\mu &= k_{2,\nu} \mathcal{S}_3^\nu 
\end{align}
\end{subequations}
 where $\mathcal{M}_3^{\mu\nu\rho}$ corresponds to the amplitude associated with the process in \cref{eq:process_3g} where the polarization vectors of the gluons have been amputated. The amputated ghost amplitudes $\mathcal{S}^\mu_i$ are defined through the Feynman diagrams in \cref{fig:ghost12,fig:ghost34,fig:ghost56} with the same index and momentum convention as in \cref{eq:process_3g} if applicable. Replacing all terms proportional to the momenta $k_1,\dots, k_3$ in the polarization sum given in \cref{eq:polSum} through the identities from \cref{eq:WardM3} for each of the three gluons and exploiting additionally the identities in \cref{eq:WardS} as well as
\begin{equation}
    (\mathcal{S}_i -\mathcal{S}_{i+1})^\ast\cdot   (\mathcal{S}_i -\mathcal{S}_{i+1}) =0, \ i=1,3,5
    \label{eq:diffSi}
\end{equation}
results in 
\begin{equation}
    - \mathcal{M}_3^{\mu\nu\rho}  \mathcal{M}_{3,\mu\nu\rho}^{\ast}+\sum_{i=1}^{6} \mathcal{S}_i^\mu \mathcal{S}_{i,\mu}^\ast
    \label{eq:M2_3g}
\end{equation}
as an expression for the squared matrix element summed over the physical final state polarizations. \Cref{eq:diffSi} follows from an explicit calculation with the help of Feynman rules. The final expression in \cref{eq:M2_3g} obeys the same structure as the one from the $2\to 2$ calculation. The ghost processes are only squared with themselves and then subtracted from the matrix element squared of the actual process.

We now turn to the discussion of the treatment of infrared divergences.
To make the integration over the three-particle phase space numerically accessible and to combine the real and virtual corrections to get an infrared safe cross section, we rely on the dipole subtraction method à la Catani-Seymour \cite{Catani:1996jh} which has recently been extended to massive initial states in the context of dark matter calculations \cite{Harz:2022ipe}. This method is based on the introduction of an auxiliary differential cross section $\dd\sigma^{\text{A}}$ which cancels the soft and collinear divergences of the differential real emission cross section pointwise but can be integrated analytically at the same time over the one-particle phase space responsible for the soft or collinear divergence. That is, the NLO correction takes the form
\begin{multline}
\Delta\sigma^{\text{NLO}}=\int_{3}\left[\dd{\sigma}^{\text{R}}_{\varepsilon=0}-\dd\sigma^{\text{A}}_{\varepsilon=0}\right]\\ +\int_{2}\left[\dd{\sigma}^{\text{V}}+\int_{1}\dd\sigma^{\text{A}}\right]_{\varepsilon=0}.
\end{multline}
According to the dipole factorization formula, the auxiliary squared matrix element related to $\dd\sigma^{\text{A}}$ for the process with three gluons in the final state consists of $27$ dipoles 
\begin{widetext}
\begin{multline}
    |\mathcal{M}_{\tilde{t}_1\tilde{t}_1^\ast \to ggg}^{\text{A}}|^2 =  \mathcal{D}_{12,3}+ \mathcal{D}_{13,2}+ \mathcal{D}_{23,1}+\mathcal{D}^{a1,b}+ \mathcal{D}^{a2,b}+ \mathcal{D}^{a3,b} + \mathcal{D}^{b1,a} + \mathcal{D}^{b2,a}+ \mathcal{D}^{b3,a}+\mathcal{D}^{a2}_1+ \mathcal{D}^{a3}_1+ \mathcal{D}^{a1}_2  + \mathcal{D}^{a3}_2\\+ \mathcal{D}^{a2}_3+ \mathcal{D}^{a1}_3+ \mathcal{D}^{b2}_1+ \mathcal{D}^{b3}_1+ \mathcal{D}^{b1}_2 + \mathcal{D}^{b3}_2+ \mathcal{D}^{b2}_3+ \mathcal{D}^{b1}_3+\mathcal{D}_{12}^a+ \mathcal{D}_{23}^a+ \mathcal{D}_{13}^a  + \mathcal{D}_{12}^b+ \mathcal{D}_{13}^b+ \mathcal{D}_{23}^b
\end{multline}
\end{widetext}
where the subscripts of the momenta in \cref{eq:process_3g} and \cref{eq:process_gqqbar} are used to label the particles. For the precise definition of the dipoles and the underlying splitting kernels we refer to Ref. \cite{Harz:2022ipe}.  For the process containing light quarks we obtain the $15$ dipoles
\begin{widetext}
\begin{equation}
    |\mathcal{M}_{\tilde{t}_1\tilde{t}_1^\ast \to \bar{q}qg}^{\text{A}}|^2 =   \mathcal{D}^{a3,b}+ \mathcal{D}^{b3,a} +\mathcal{D}^{a3}_1+ \mathcal{D}^{a3}_2+ \mathcal{D}^{b3}_1+ \mathcal{D}^{b3}_2  +\mathcal{D}_{12,3}+ \mathcal{D}_{13,2}+ \mathcal{D}_{23,1} +\mathcal{D}_{12}^a+ \mathcal{D}_{12}^b+ \mathcal{D}_{31}^a  + \mathcal{D}_{32}^a+ \mathcal{D}_{31}^b+ \mathcal{D}_{32}^b.
\end{equation}
\end{widetext}
For the explicit construction of the insertion operator which cancels the infrared divergences on the virtual side, we refer again to Ref. \cite{Harz:2022ipe} due to the large number of terms coming from the non-factorizable color and spin structures.


\subsection{\label{sec:sommer}Sommerfeld enhancement}
We have discussed the fixed-order NLO corrections in the previous two subsections. However, for the non-relativistic regime, as it is typical during freeze-out, there are also important contributions to the relic density from the exchange of $n$ potential gluons between the incoming stop and antistop giving a correction factor proportional to $(\alpha_s/v)^n$. This is the well-known Sommerfeld enhancement \cite{Sommerfeld:1931qaf} of higher-order terms which can spoil the perturbativity of the cross section when the relative velocity is of the order of the strong coupling, and therefore these terms need to be resummed to all orders in perturbation theory. 
The fact that the tree-level cross section is dominated by $S$-wave annihilation as discussed in \cref{sec:scenario} and visible in \cref{fig:LO-XSec}, allows to compute the Sommerfeld enhanced cross section
\begin{multline}
\label{eq:sigmares}
    (\sigma v)^{\text{Som}} = S_{0,[\mathbf{8}]}\left((\sigma v)^{\text{Tree}}_{ gg,[\mathbf{8_S}]} + (\sigma v)^{\text{Tree}}_{gg,[\mathbf{8_A}]} \right. \\
    \left. + N_f (\sigma v)^{\text{Tree}}_{q\bar{q},[\mathbf{8}]} \right) +S_{0,[\mathbf{1}]}\,(\sigma v)^{\text{Tree}}_{gg,[\mathbf{1}]} 
\end{multline}
by multiplying the leading contribution with the Sommerfeld factor 
\begin{align}
\label{eq:Sommerfeldfactor}
	S_{0,\bf{[R]}} = \frac{\Im\mathcal{G}^{\bf{[R]}}(\vec{r}=0,\sqrt{s}+i\Gamma_{\tilde{t}_1})}{\Im\mathcal{G}_0(\vec{r}=0,\sqrt{s}+i\Gamma_{\tilde{t}_1})}\,,
\end{align} 
whose computation follows the standard framework of non-relativstic QCD (NRQCD) described in Refs.  \cite{Hagiwara:2008df,Kiyo:2008bv}. The Green's function $\mathcal{G}^{\bf{[R]}}(\vec{r}=0,\sqrt{s}+i\Gamma_{\tilde{t}_1})$ is defined as solution of the Schrödinger equation
\begin{align}
\label{eq:SchroedingerEQ}
	\Big[ H^{[\mathbf{R}]} - \big( \sqrt{s}+i\Gamma_{\tilde{t}_1} \big) \Big]
		\mathcal{G}^{[\bf R]} \big( \vec{r};\sqrt{s}+i\Gamma_{\tilde{t}_1}\big)\nonumber 
		= \delta^{(3)}(\vec{r})
\end{align} 
evaluated at the origin where
\begin{equation}
    H^{[\mathbf{R}]} = 2 m^2_{\tilde{t}_1}-\frac{1}{m_{\tilde{t}_1}} \laplacian  +V^{[\bf{R}]}(\vec{r}).
\end{equation}
is the Hamiltonian of the quasi-stoponium. 
The corresponding Coulomb QCD potential receives important contributions from gluon and fermion loops and reads at NLO in momentum space
\begin{multline}
\label{eq:Coulombpotential}
	\tilde{V}^{[\bf{R}]}(\vec{q}) = -C^{[\bf{R}]}\,\frac{4 \pi \alpha_{s}(\mu_C)}{\vec{q}^2} \\  \times \bigg\{ 1 + \frac{\alpha_s(\mu_C)}{4\pi} \bigg[\beta_0\ln \left(\frac{\mu^2_C}{\vec{q}^2}\right) +a_1 \bigg]  \bigg\}\,
\end{multline}
with the color factors 
\begin{align}
    C^{[\mathbf{1}]}=C_F, \ \ \ \ \ 
    C^{[\mathbf{8}]}=C^{[\mathbf{8}_S]}=C^{[\mathbf{8}_A]}=-\frac{1}{2 N_c}
\end{align}
and the constants
\begin{align}
    &a_1 = \frac{31}{9} C_A -\frac{20}{9} T_F n_f \\
    &\beta_0 = \frac{11}{3}C_A - \frac{4}{3}T_f n_f
\end{align}
where we work with $n_f=5$.
The analytic solution for the Green's function at the origin at NLO accuracy is 
\begin{align}
	G^{\bf{[R]}}({\vec{r}=0};\sqrt{s}+i\Gamma_{\tilde{t}_1})&=  \frac{C^{\bf{[R]}}\alpha_{s}(\mu_C) m^2_{\tilde{t}_1}} {4\pi}\nonumber \\
	&\times \Big[ g_{\mathrm{LO}} + \frac{\alpha_{s}(\mu_C)}{4\pi}g_{\mathrm{NLO}} \Big]\,,
\end{align}
where the LO and NLO contributions are 
\begin{align}\label{eq:glo}
	&g_{\mathrm{LO}}\hspace{2mm} = -\frac{1}{2\kappa}+L - \psi^{(0)}, \\ \label{eq:gnlo}
	&g_{\mathrm{NLO}}= 	\beta_0 \Big[ L^2 - 2L(\psi^{(0)} - \kappa\psi^{(1)}) + \kappa\psi^{(2)} + 
		(\psi^{(0)})^2  \nonumber \\
		&-3\psi^{(1)} - 2\kappa\psi^{(0)}\psi^{(1)} 
		+ 4 \ _4F_3(1,1,1,1;2,2,1-\kappa;1)\Big] \nonumber \\
	& + a_1 \Big[L-\psi^{(0)}+\kappa\psi^{(1)} \Big]\,.
\end{align}
Here, the constants 
\begin{align}
	\kappa&=\frac{i C^{\bf{[R]}}\alpha_{s}(\mu_C)}{2v_s}, \\ 
	L & =  \ln\frac{i \mu_C}{2 m_{\tilde{t}_1 }v_{s}}
\end{align}
are defined through the non-relativistic velocity of the incoming particles 
\begin{equation}
    v_{s}  =  \sqrt{\frac{\sqrt{s}+i\Gamma_{\tilde{t}_1}-2m_{\tilde{t}_1}}{ m_{\tilde{t}_1}}} 
\end{equation}
and $\psi^{(n)}=\psi^{(n)}(1-\kappa)$ is the $n$-th derivative of $\psi(z) = \gamma_{\mathrm{E}} + \mathrm{d}/\mathrm{d}z\ln\Gamma(z)$ with the argument $(1 - \kappa)$. 
For the computation of the Sommerfeld factor, we also need the free Greens's function 
\begin{align}
    \mathcal{G}_0(0,\sqrt{s}+i\Gamma_{\tilde{t}_1}) = \frac{ i m^2_{\tilde{t}_1} v_{\rm s}}{4\pi}\,.
\end{align}
We address now the choice for the Coulomb scale $\mu_C$ at which the strong coupling in the QCD potential is evaluated. Following Ref. \cite{Beneke:2010da}, we set 
\begin{align}
    \label{eq:Coulombscale}
    \mu_C = \mathrm{max} \left\{ 2 m_{\tilde{t}_1} v_{s}, \mu_B \right\}\,,
\end{align}
where $2 m_{\tilde{t}_1} v_{s}$ is motivated by the typical momentum transfer mediated by the potential gluons. The Bohr scale $\mu_B$ corresponds to twice the inverse Bohr radius $r_B$ and is obtained by iteratively solving the equation
\begin{align}
\mu_B \equiv 2/r_B = C_F m_{\tilde{t}_1} \alpha_s(\mu_B).
\end{align}
 For the scenario in \cref{tab:scenario}, the Bohr scale takes the value $\mu_B  = \SI{204}{\giga\electronvolt}$ and the associated value for the strong coupling in the $\overline{\text{MS}}$-scheme with $6$ active quark flavors is $\alpha_s(\mu_B)=\SI{0.1058}{}$.
 
 As a single gluon exchange is already included in our fixed-order NLO calculation (see \cref{fig:diags_squark-gluon-vertex} and \cref{fig:four-vertex}), we have to match it to the Sommerfeld enhanced cross section in order to avoid double counting. This is achieved by taking only the terms of $\order{\alpha^2_s}$ in \cref{eq:Sommerfeldfactor} into account giving the full cross section $(\sigma v)^{\text{Full}}$.
 
 As described in Ref. \cite{Schmiemann:2019czm}, it is also possible to subtract the velocity-enhanced part from the fixed order calculation in order to obtain the "pure" NLO cross section which gives
\begin{multline}
    (\sigma v)_v^{\text{NLO}} =  (\sigma v)^{\text{NLO}} + \frac{\alpha_s(\mu_R) \pi}{v_{\text{rel}}} \\ \times \left( \sum_{\mathbf{R}} C^{[\mathbf{R}]} (\sigma v)^{\text{Tree}}_{ gg,[\mathbf{R}]} + N_f C^{[\mathbf{8}]} (\sigma v)^{\text{Tree}}_{q\bar{q}} \right)
\end{multline}
with the relativistic relative velocity 
\begin{equation}
    v_{\text{rel}} =\frac{ v }{2-\rho} .
\end{equation}

\section{\label{sec:results}Numerical results}
In this section, we discuss the impact of the corrections on the stop-antistop annihilation cross section and the corresponding impact on the theoretical uncertainty deduced from scale variations. Then, we study the impact of the full correction on the relic density for stop-antistop annihilation alone as well in conjunction with the other two important processes shown in \cref{tab:channels}. 
\subsection{Annihilation cross section and its theoretical uncertainty}
\begin{figure*}[ht]
    \centering
    \begin{subfigure}[c]{0.55\textwidth}		
    \centering	
    \includegraphics[width=\textwidth]{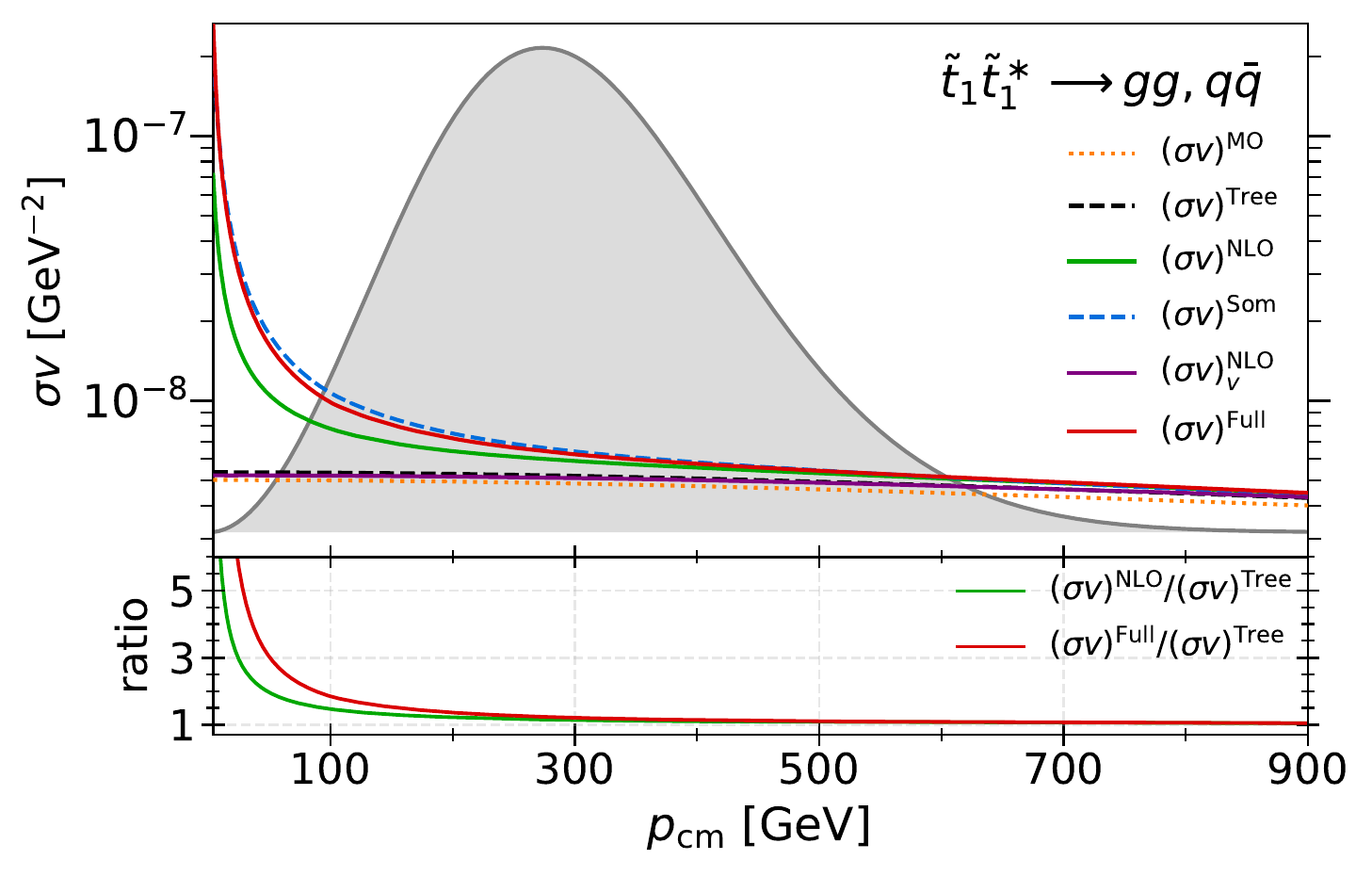}
    \subcaption{Stop-antistop annihilation.}
    \label{fig:XSec-NLO-stsT2xx}
    \end{subfigure}
    
    \begin{subfigure}[c]{0.45\textwidth}
    \includegraphics[width=\textwidth]{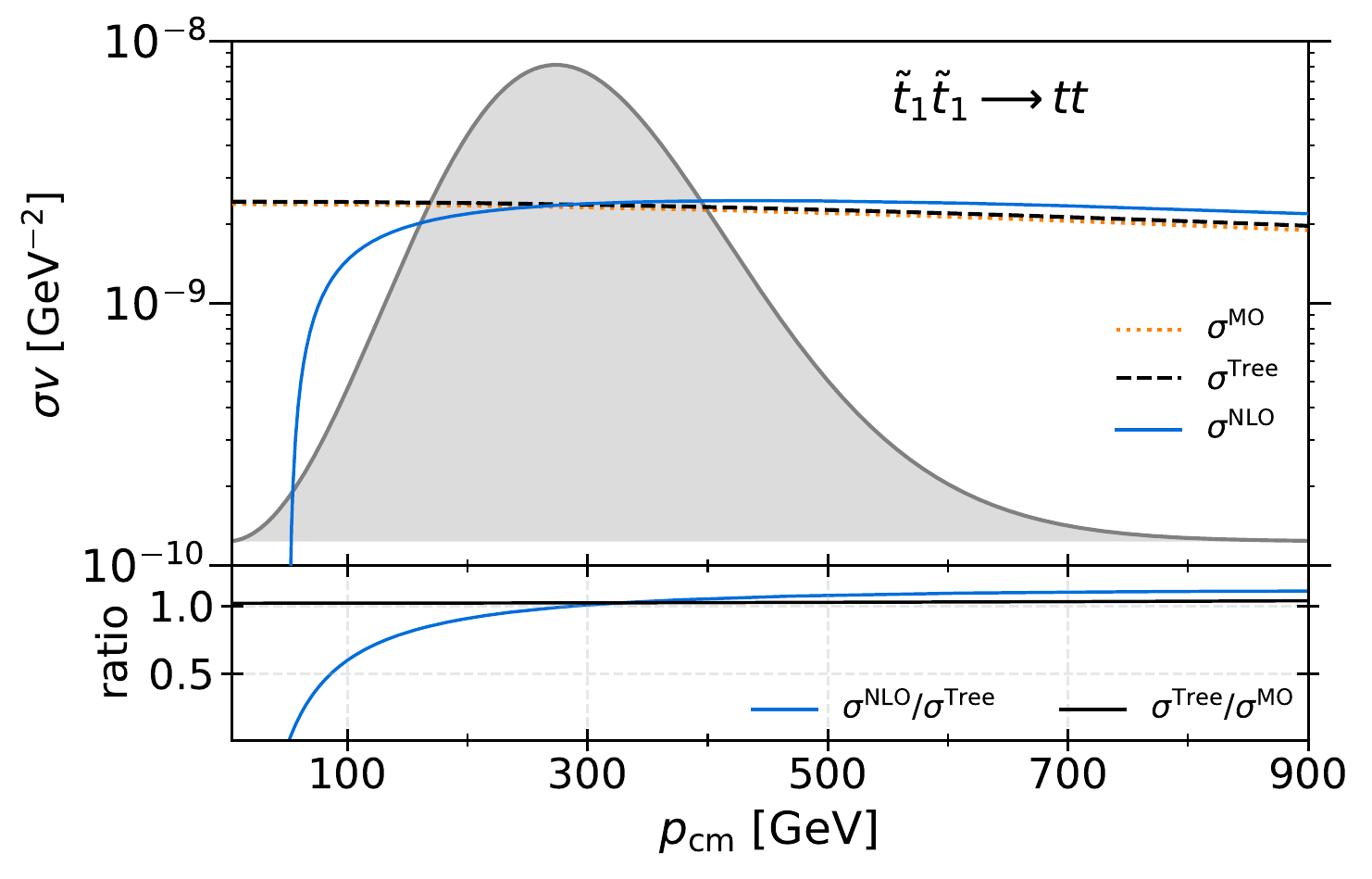}
    \subcaption{Stop pair-annihilation.}
    \label{fig:XSec-NLO-stst2QQ}
    \end{subfigure}
    \begin{subfigure}[c]{0.45\textwidth}
    \includegraphics[width=.95\textwidth]{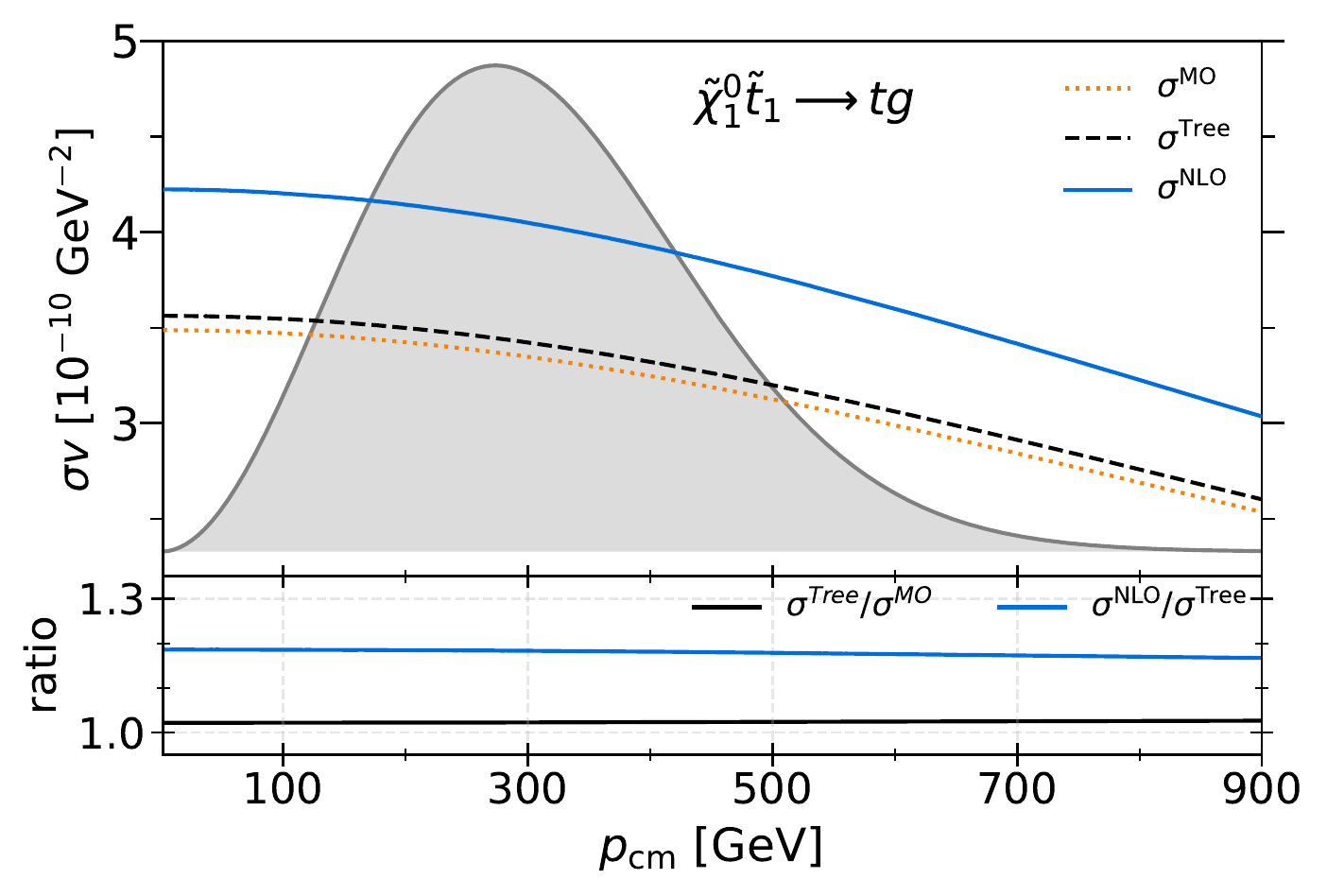}
    \subcaption{Neutralino-stop coannihilation.}
    \label{fig:XSec-NLO-NeuQ2qg}
    \end{subfigure}
    \caption{Annihilation cross section $\sigma v$ for stop-antistop annihilation into light quarks and gluons, stop annihlation into top quarks and neutralino-stop coannihilation into a top and a gluon. The lower parts of the plots show ratios of cross sections.}
    \label{fig:NLO-QQ-qG}
\end{figure*}
In \cref{fig:XSec-NLO-stsT2xx}, we show the stop-antistop annihilation cross cross section as a function of the CM momentum $p_{\text{cm}}$ for the parameter point defined in \cref{tab:scenario}. More precisely, we show the cross section at tree-level as provided by \texttt{DM@NLO} (black dashed line) and by \texttt{MicrOMEGAs 2.4.1} (dotted orange line), including the NLO corrections (green solid line) and the full cross section with the Sommerfeld enhancement effect (red solid line). In addition, we show the pure Sommerfeld enhanced cross section (blue dashed line) and the "pure" NLO cross section without the velocity-enhanced part (purple solid line). For small relative velocities the Coulomb corrections from the exchange of multiple gluons between the incoming particles dominate the full corrected annihilation cross section. As discussed in \cref{sec:sommer}, the effect of the Coulomb corrections depends on the quadratic Casimir of the representation under which the incoming particles transform. The singlet feels an attractive force whereas the squark and antisquark transforming under an eight dimensional representation are repelled from each other. In this case, the Coulomb corrections increase the annihilation probability so that the full corrected cross section becomes larger than $\SI{100}{\percent}$ of the tree-level cross section for CM momenta below $\SI{88}{\giga\electronvolt}$ even though the LO cross section is dominated by the symmetric octet contribution which is due to the color suppression given by $\nicefrac{1}{2 N_c}$ in the Sommerfeld factor for the eight dimensional representation.  For vanishing relative velocities, the enhanced cross section even diverges and approaches the well-known Coulomb singularity which could be cured by taking the formation of bound states into account properly. However, as the Boltzmann distribution almost vanishes for momenta around $p_{\text{cm}}=0$, such effects are heavily suppressed. In contrast, the "pure" NLO correction without any enhancement corresponds to an improvement of less than $\pm\SI{3}{\percent}$ of the LO cross section such that the full corrected cross section is in very good approximation given by the pure Sommerfeld enhancement, i.e. $(\sigma v)^{\text{Full}}\approx (\sigma v)^{\text{Som}}$. 

The other two processes which we include in our analysis and are important in the region around the reference scenario, namely $\tilde{t}_1 \tilde{t}_1 \to tt$ and $\tilde{\chi}^0_1 \tilde{t}_1 \to t g$, have been investigated in the context of \texttt{DM@NLO} in Refs. \cite{Schmiemann:2019czm,Harz:2014tma}. In contrast to the two original publications, we do not use the phase space slicing method for the real corrections in this paper but the dipole subtraction method. The implementation of the dipole approach for the two processes and the comparison between both methods were the subjects of Ref. \cite{Harz:2022ipe}. The corresponding tree-level cross sections obtained with \texttt{MicrOMEGAs 2.4.1} (orange dotted line) and with \texttt{DM@NLO} (black dashed line) including the NLO corrections (blue solid line) are shown for both channels in \cref{fig:XSec-NLO-stst2QQ,fig:XSec-NLO-NeuQ2qg}, respectively. Even though \texttt{MicrOMEGAs} uses an effective top quark mass which evaluates at the scale $\mu_{\text{MO}}$ to $m_t^{\text{eff}}=\SI{146.2}{\giga\electronvolt}$ instead of the corresponding on-shell value $m_t=\SI{173.2}{\giga\electronvolt}$ which is used by \texttt{DM@NLO}, the difference between the two tree-level cross sections in \cref{fig:XSec-NLO-NeuQ2qg} is due to the differences in the strong coupling as discussed in the context of the LO cross section of $\tilde{t}_1 \tilde{t}_1^\ast \to gg$.  In the case of stop pair-annihilation, the NLO corrections cause a positive shift of about $\SI{10}{\percent}$ for large $p_{\text{cm}}$ around $\SI{600}{\giga\electronvolt}$ compared to the tree-level cross section whereas the correction becomes large and negative for CM momenta less than $\SI{287}{\giga\electronvolt}$. For CM momenta below $\SI{50}{\giga\electronvolt}$ the total cross section becomes negative which is unphysical but we make in the following the assumption that this momentum region is irrelevant for the computation of the relic density due to an almost vanishing Boltzmann distribution for such low velocities. Furthermore, this unphysical behavior has already been extensively discussed in Ref. \cite{Schmiemann:2019czm}.
In the case of neutralino-stop coannihilation the correction is stable around $\SI{19}{\percent}$ for all relevant CM momenta. 

We continue with the analysis of the theoretical uncertainties of the stop-antistop annihilation cross section from variations of the Coulomb and renormalization scale where we identify the central scales with the ones used in the previous discussion, i.e. $\mu_R^{\text{central}} = Q_{\text{SUSY}}$ and $\mu^{\text{central}}_C =  \mathrm{max} \left\{ 2 m_{\tilde{t}_1} v_{s}, \mu_B \right\}$. 
\begin{figure*}[ht]
    \centering
    \includegraphics[width=0.6\textwidth]{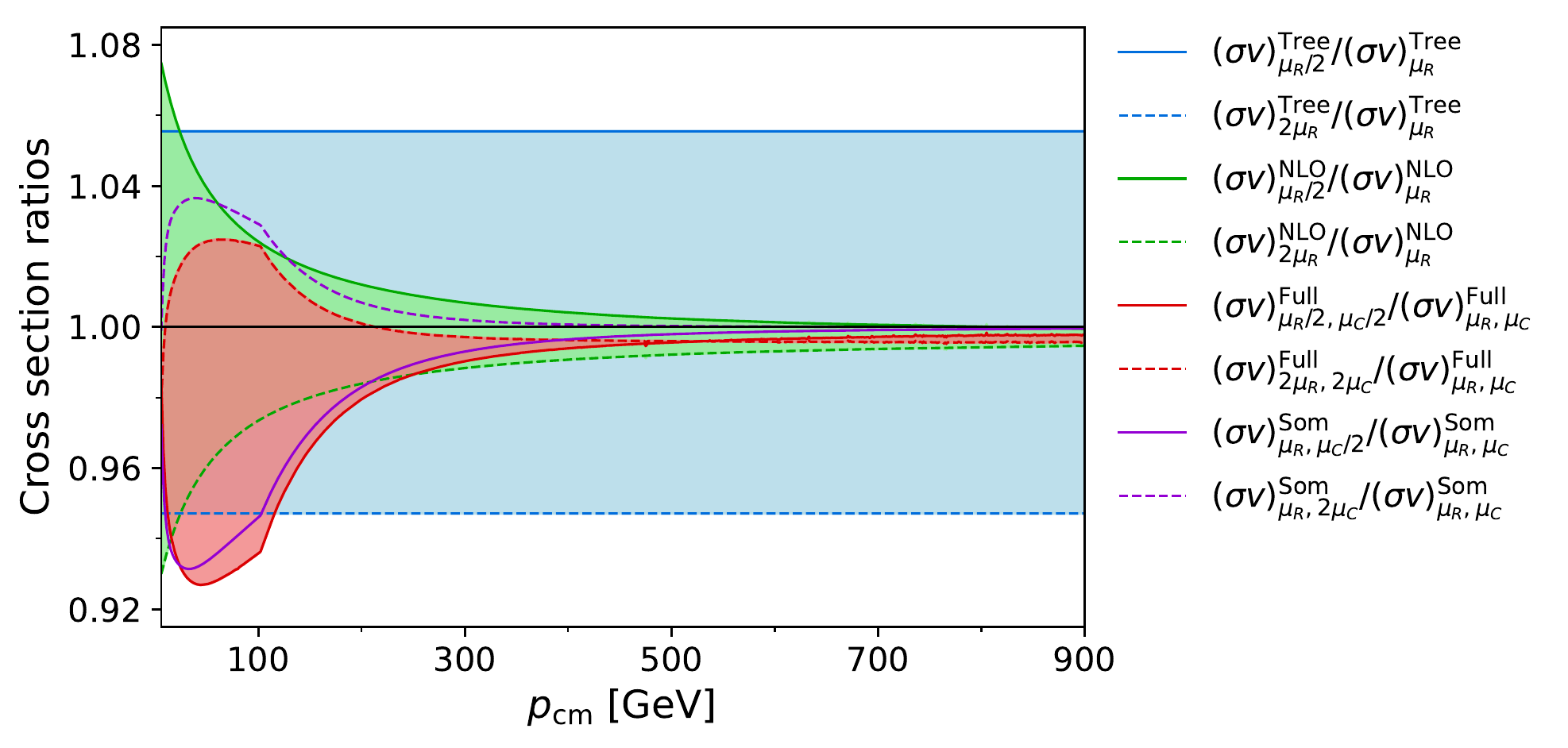}
    \caption{Cross section of the stop-antistop annihilation process for different values of the renormalization scale (and Coulomb scale) in dependence of the CM momentum and normalized to the cross section obtained at the central scale(s).}
    \label{fig:XSec_pcm_scale}
\end{figure*}
\begin{figure*}[ht]
    \centering
    \includegraphics[width=0.6\textwidth]{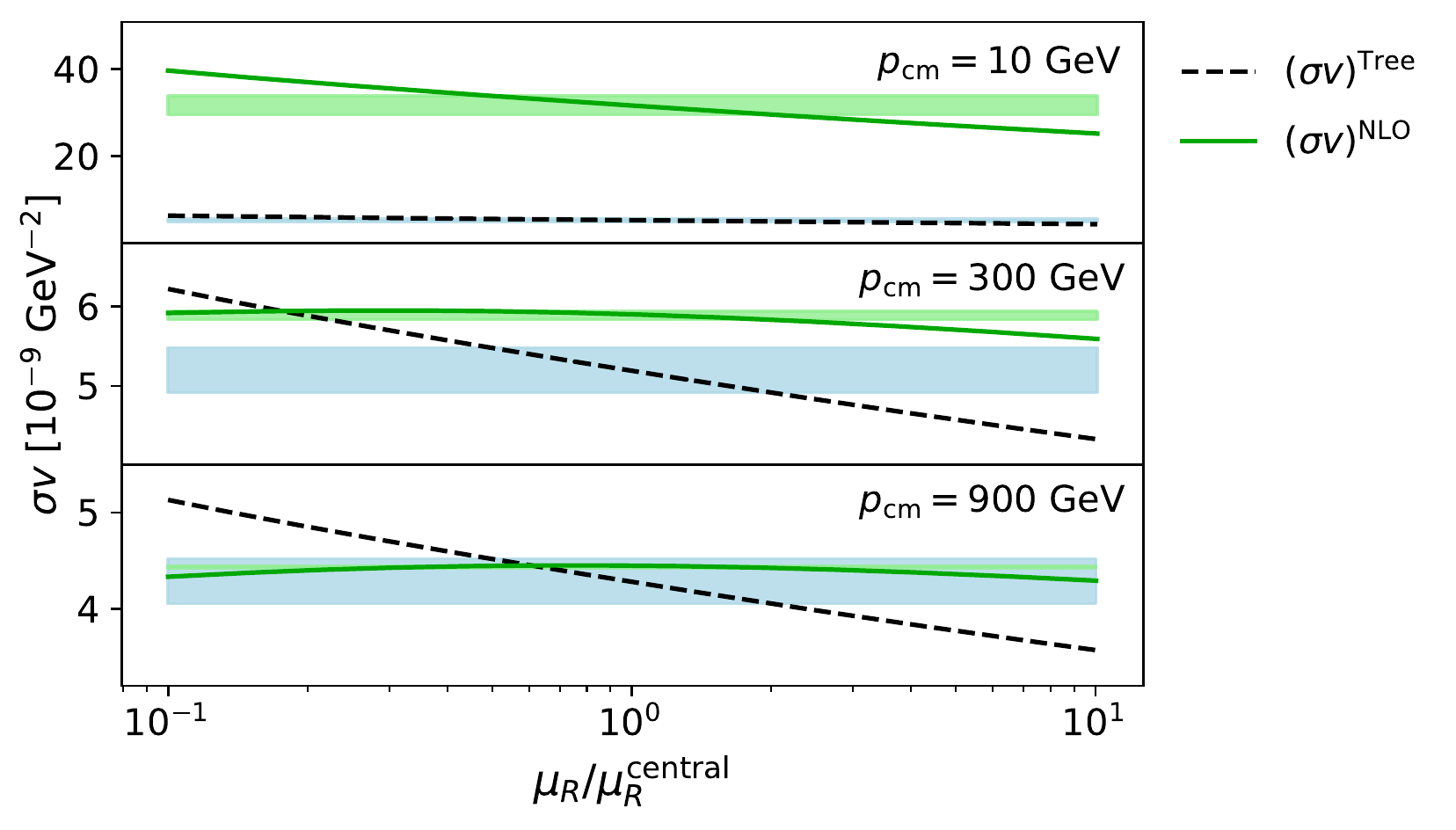}
    \caption{Renormalization scale dependence of the LO and NLO cross section corresponding to stop-antistop annihilation into gluons and light quarks for three different CM momenta. The colored bands indicate the scale variation of \cref{fig:XSec_pcm_scale}.}
    \label{fig:XSec_mu_scale}
\end{figure*}
In \cref{fig:XSec_pcm_scale}, we vary $\mu_R$ and $\mu_C$ by factors of two and show the associated values of the annihilation cross section at tree-level (blue), at NLO (green) including the Coulomb corrections (red) as well as the pure Sommerfeld enhanced cross section (purple) normalized to the corresponding cross section obtained at the central scale(s). In conjunction, the LO and NLO cross section as function of the renormalization scale for three different CM momenta are shown in \cref{fig:XSec_mu_scale}. Within the chosen renormalization scheme, the scale dependence enters the tree-level cross section only through the strong coupling and we estimate the theoretical uncertainty to about $\pm\SI{5.5}{\percent}$. For large CM momenta ($p_{\text{cm}}\approx \SI{900}{\giga\electronvolt}$) the NLO correction lies within the LO uncertainty and the theoretical uncertainty is reduced to below $\SI{1}{\percent}$. For intermediate energies ($p_{\text{cm}}\approx \SI{300}{\giga\electronvolt}$) the NLO correction is no longer contained in the LO uncertainty but the uncertainty is still reduced to about $\pm\SI{1.5}{\percent}$ by including the higher-order corrections. For very small relative velocities ($p_{\text{cm}}\approx \SI{10}{\giga\electronvolt}$) the cross section becomes non-perturbative and the NLO uncertainty is larger than the LO one. However, by including the Coulomb corrections the upper uncertainty bound for small energies is halved whereas the lower uncertainty bound increases and we have only a reduction for $v\to 0$. As the full corrected cross section is in very good approximation given by the Sommerfeld enhancement only, we expect the same for the associated uncertainty which turns out to be the case. We note at this point that the kink in the uncertainty band of $(\sigma v)^{\text{Full}}$ and $(\sigma v)^{\text{Som}}$ comes from the transition from the Bohr scale to the scale of the typical momentum exchange $2 m_{\tilde{t}_1} v_s$.

\subsection{Impact on the relic density}
\begin{figure*}[ht]
    \begin{subfigure}[c]{0.45\textwidth}
    \includegraphics[width=\textwidth]{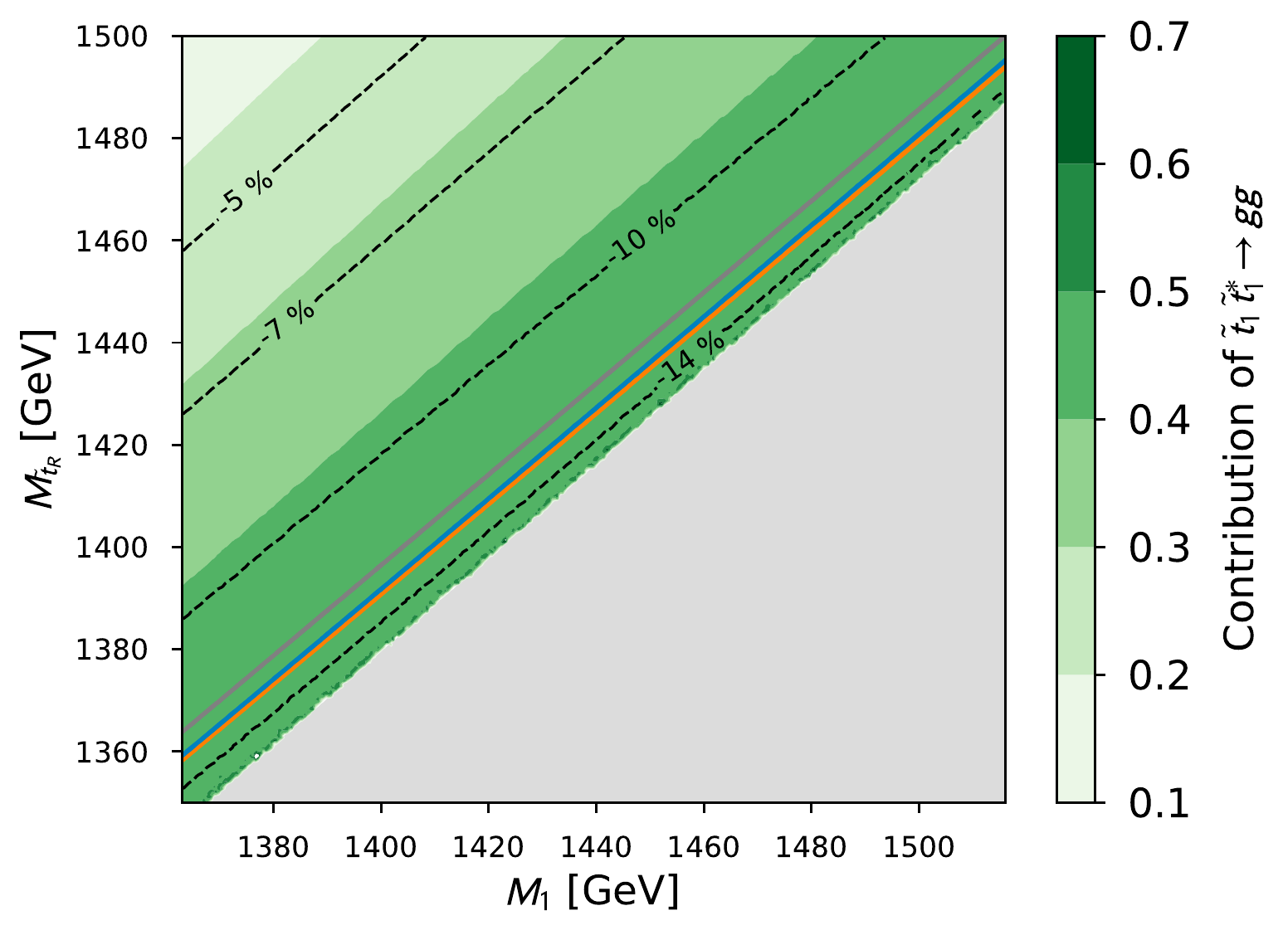}
    \subcaption{Only $\tilde{t}_1 \tilde{t}_1^\ast \to gg,q\overline{q}$.}
    \label{fig:relic-NLO-stsT2xx}
    \end{subfigure}
    \begin{subfigure}[c]{0.45\textwidth}
    \includegraphics[width=\textwidth]{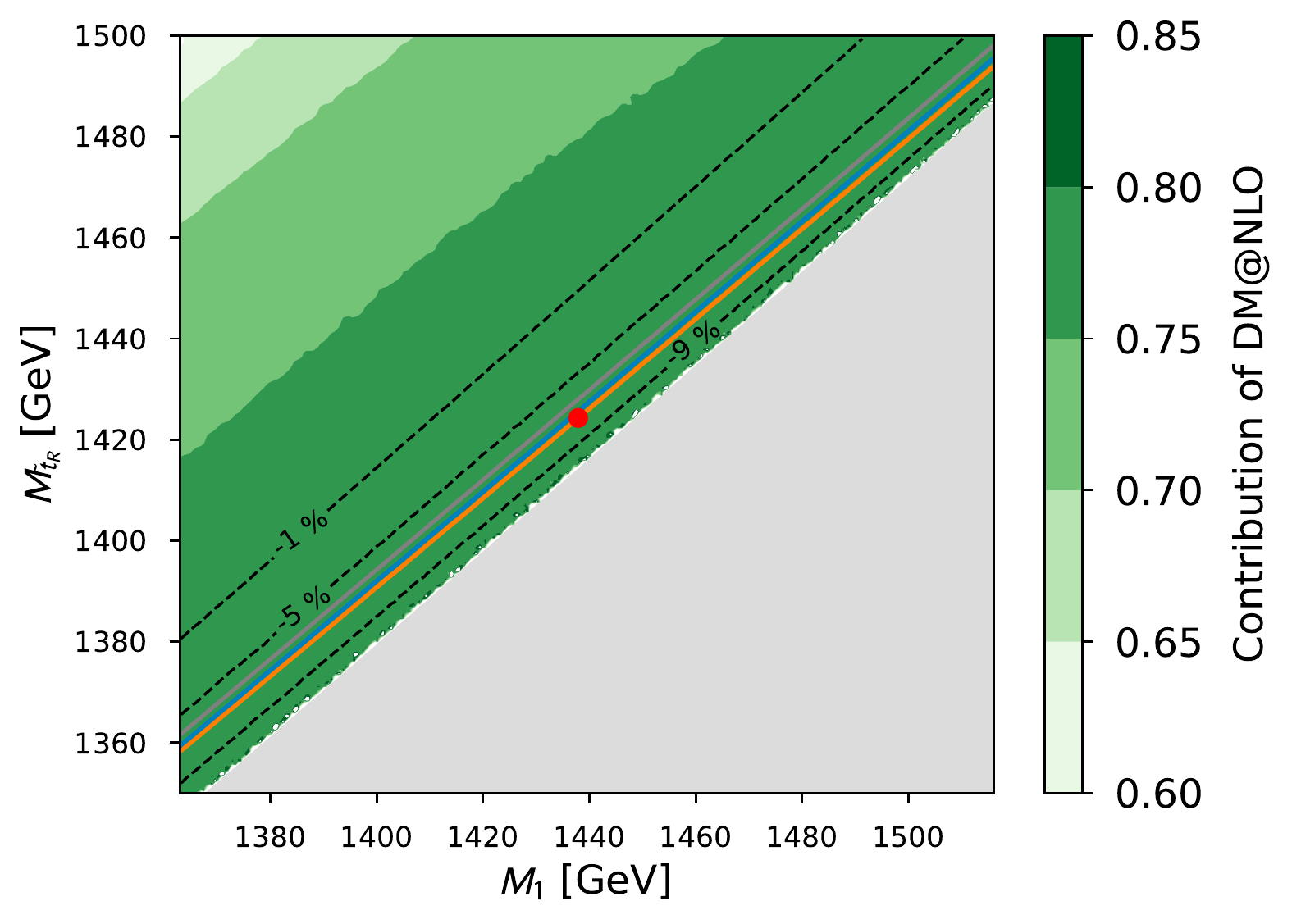}
    \subcaption{\texttt{DM@NLO} total.}
    \label{fig:relic-NLO-dmnlo-total}
    \end{subfigure}
    \caption{Parameter region in the $M_1$-$M_{\tilde{t}_R}$ plane that is consistent with the \emph{Planck} limit in \cref{eq:omh2} at the $2\sigma$ confidence level. The orange band corresponds to the \texttt{MicrOMEGAs 2.4.1} calculation, the blue one to the \texttt{DM@NLO} tree-level cross section and the gray band to the full corrected cross section. In the right panel, all three important processes are included whereas only the stop-antistop annihilation cross section is replaced by \texttt{DM@NLO} in the left panel. The shades of green indicate the total contribution of corrected (co)annihilation processes to the relic density and the black solid lines the relative change in the relic density compared to our tree-level result. }
    \label{fig:relic-NLO}
\end{figure*}
At last, we investigate the impact of our radiative corrections on the neutralino relic density $\Omega_{\tilde{\chi}^0_1} h^2$ by including all three processes from \cref{tab:channels} which are important in a region around the chosen reference scenario and are available in \texttt{DM@NLO} as well as for the process which is subject of this paper only. This means that the integration of the Boltzmann equation in \cref{eq:boltzmann} is still performed by \texttt{MicrOMEGAs 2.4.1} but the cross sections are replaced by the ones implemented in \texttt{DM@NLO} for the specified cases and still obtained from $\texttt{CalcHEP}$ for the remaining ones. Similar to \cref{sec:pheno}, we study the impact on the relic density in the plane spanned by $M_1$ and $M_{\tilde{t}_R}$ which is shown for both cases in \cref{fig:relic-NLO}. 

\begin{figure*}[htb]
    \begin{subfigure}[c]{0.45\textwidth}
    \includegraphics[width=\textwidth]{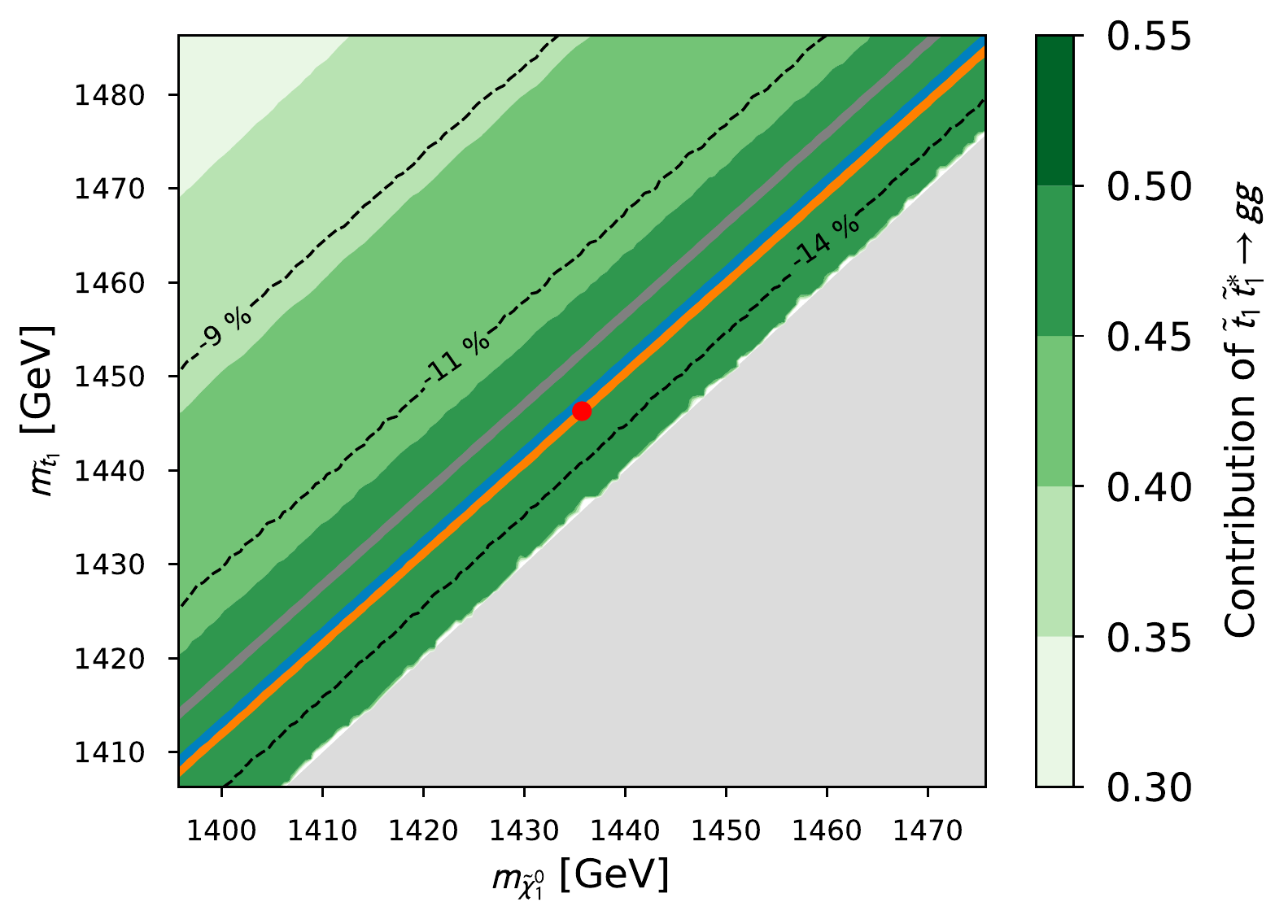}
    \end{subfigure}
    \begin{subfigure}[c]{0.45\textwidth}
    \includegraphics[width=\textwidth]{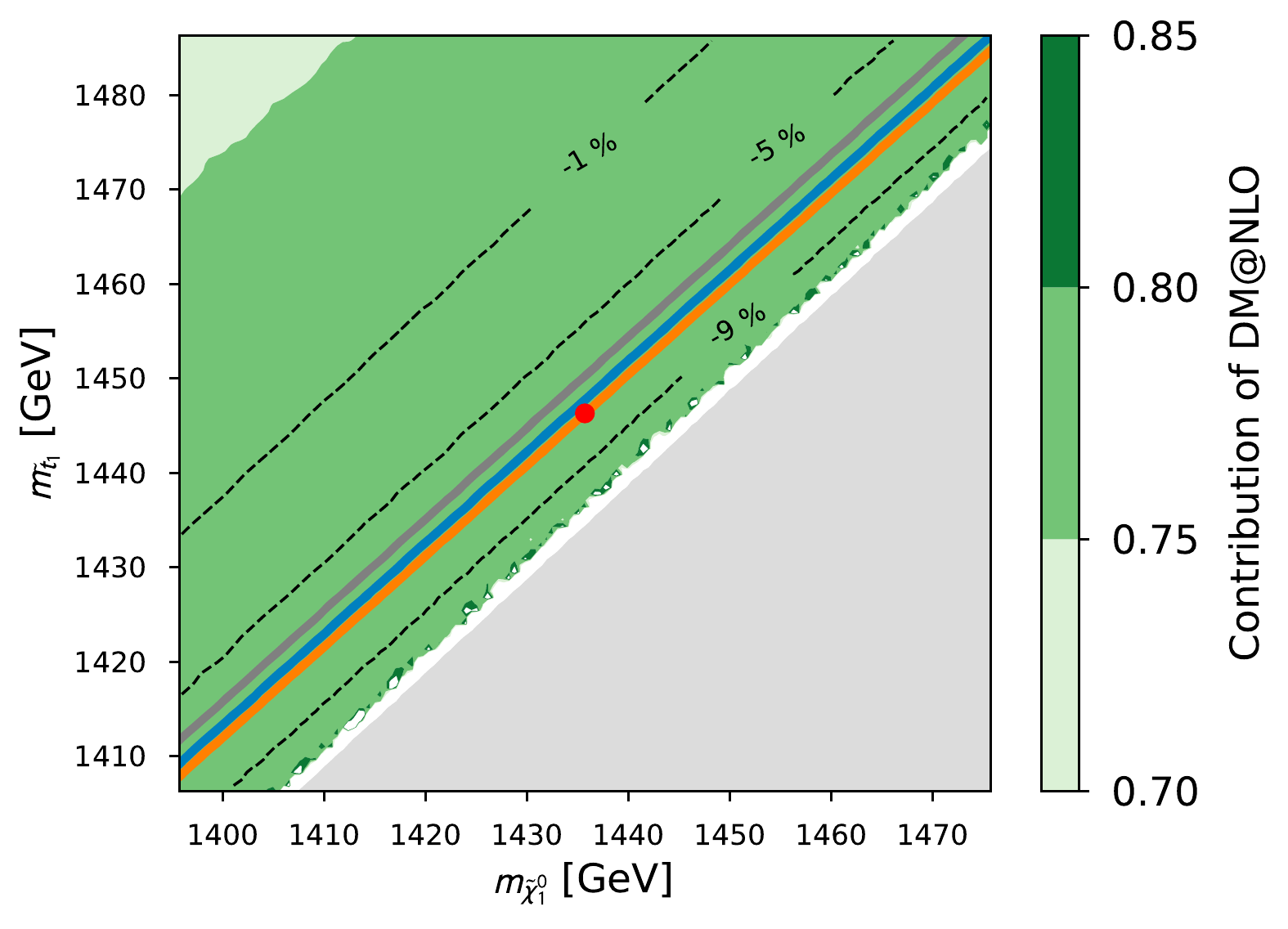}
    \end{subfigure}
    \caption{Same as \cref{fig:relic-NLO}, but projected into the plane of the physical neutralino and stop masses and enlarged.}
    \label{fig:relic-zoom}
\end{figure*}

As before, the region which is compatible up to two sigma with the \emph{Planck} limit is shown in orange for the values obtained with \texttt{MicrOMEGAs 2.4.1}, in blue for the tree-level values from \texttt{DM@NLO} and in gray for the radiative corrections. In addition, the same results are presented in \cref{fig:relic-zoom} projected into the plane of the physical neutralino and stop mass where one should highlight that this variation only comes from the scan over the parameters $M_1$ and $M_{\tilde{t}_R}$ whereas all other parameters in \cref{tab:scenario} remain fixed. The small difference between the tree-level results is again mainly due to the differences in the strong coupling.

In both cases, the favored parameter region consistent with the \emph{Planck} limit is shifted towards larger stop masses for a fixed neutralino mass to compensate the increased effective annihilation cross section where this shift exceeds the experimental uncertainty. However, if we only include the radiative corrections for stop-antistop annihilation the cosmologically favored stop mass is increased by about $\SI{6.1}{\giga\electronvolt}$ compared to the \texttt{MicrOMEGAs} result whereas the additional inclusion of the higher-order corrections to the processes $\tilde{\chi}^0_1 \tilde{t}_1 \to t g$ and $\tilde{t}_1 \tilde{t}_1 \to t t$ reduces this shift to about $\SI{4.3}{\giga\electronvolt}$. This is due to the large negative NLO corrections for small $p_{\text{cm}}$ that occur for stop pair-annihilation.  

\section{\label{sec:conclusion}Conclusion}
The annihilation of colored particles which are close in mass to the dark matter candidate is an important mechanism to allow for higher dark matter masses while still being able to explain the measured relic density. In the MSSM, a theoretically well motivated candidate for such annihilation processes is the lightest stop. Based on previous analyses which show that the inclusion of higher-order corrections to the relic density exceeds the experimental uncertainty of the dark matter content in the universe, we have presented in this paper NLO SUSY-QCD corrections to stop-antistop annihilation into gluons and light quarks including QCD Coulomb corrections of $\order{\alpha^2_s}$. The two processes $\tilde{t}_1 \tilde{t}_1^\ast \to g g$ and $\tilde{t}_1 \tilde{t}_1^\ast \to q \overline{q}$ with $q$ being an effectively massless quark are combined in our analysis since we found within our calculation that these two processes can not be treated separately at NLO accuracy in order to guarantee a well-defined and infrared safe cross section. 
In order to study the impact of such corrections on the annihilation cross section itself and the relic density, we have performed a random scan in the phenomenological MSSM with 19 free parameters to select a reference scenario that is consistent with the current most import experimental constraints and contains a stop with almost the same mass as the neutralino. 
The numerical analysis showed that the resummed cross section matched to the fixed-order NLO calculation is in very good approximation given by the Sommerfeld enhanced cross section only which can in turn be used to significantly speed up relic density scans while capturing the majority of the NLO corrections. We are confident that this result extends to simplified dark matter models containing a colored scalar similar to the MSSM as those proposed for LHC searches in Ref. \cite{Abdallah:2015ter}. 
In addition, we observed that the inclusion of the NLO corrections reduces the dependence of the cross section on the renormalization scale in the perturbative regime from $\pm\SI{5.5}{\percent}$ to below $\pm\SI{2}{\percent}$.
Finally, we found with respect to the impact on the relic density that the corrections to stop-antistop annihilation only can shift the cosmologically favored parameter region by a few GeV and they are therefore larger than the current experimental uncertainty. However, through the additional inclusion of the NLO SUSY-QCD corrections to $\tilde{\chi}^0_1 \tilde{t}_1 \to t g$  and $\tilde{t}_1 \tilde{t}_1 \to tt$ this shift is reduced by about $\SI{30}{\percent}$ due to a large negative correction for the stop pair-annihilation.
As in our previous studies, we conclude that the identification of parameter regions consistent with the measured relic density at the current level of precision requires the inclusion of NLO and Coulomb corrections including those covered in this work.

\begin{acknowledgments}
M.K. thanks the School of Physics at the University of New South Wales in Sydney, Australia for its hospitality and financial support through the Gordon Godfrey visitors program. The work of M.K. was also funded by the DFG through grant KL 1266/10-1, the work by M.K., K.K. L.P.W. through the DFG Research Training Group 2149 "Strong and Weak Interactions - from Hadrons to Dark Matter". The figures and Feynman diagrams presented in this paper have been generated using \texttt{MatPlotLib} \cite{Hunter:2007} and \texttt{TikZ-Feynman} \cite{Ellis:2016jkw}.
\end{acknowledgments}



%

\end{document}